\documentclass[epj]{svjour}

\usepackage[utf8]{inputenc}
\usepackage[T1]{fontenc}
\usepackage{eurosym}
\usepackage{amsmath,amssymb,bm}
\usepackage{mathtools}
\usepackage{graphics}
\usepackage{graphicx}
\usepackage{scalerel}
\usepackage{tabularx}
\usepackage{multirow}
\usepackage{psfrag}
\usepackage{comment}
\usepackage{pifont}
\usepackage{breqn}
\usepackage{url}
\usepackage{epsfig}
\usepackage{array}
\usepackage{epstopdf}
\usepackage{overpic}
\usepackage{xspace}
\usepackage[english]{babel}
\usepackage{cite}
\usepackage{color}
\usepackage{wasysym}

\usepackage{siunitx,booktabs}
\newcommand{\ra}[1]{\renewcommand{\arraystretch}{#1}}

\newcommand{\orcid}[1]{\href{https://orcid.org/#1}{\includesgraphics[height = 1.5ex]{fig/ORCID_iD.png}}}
\newcommand\orcidicon[1]{\href{https://orcid.org/#1}{\mbox{\scalerel*{
\includesgraphics{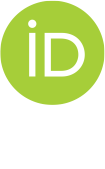}
}{\'E}}}}

\newcommand{\be}{\begin{equation}}
\newcommand{\ee}{\end{equation}}
\newcommand{\ba}{\begin{eqnarray}}
\newcommand{\ea}{\end{eqnarray}}
\newcommand{\morat}{{\textsc{morat}}rap\xspace}
\newcommand{\lpct}{{\textsc{lpct}}rap\xspace}

\usepackage{hyperref}

\begin{document}
%
%
\title{Geometry optimisation of a transparent axisymmetric ion trap for the MORA project}

\author{
M.~Benali\inst{1}\fnmsep\thanks{\email{benali@lpccaen.in2p3.fr}}\fnmsep\href{https://orcid.org/0000-0001-6503-8040}{\includegraphics[height = 2.2ex]{fig/ORCIDiD_exponent.png}}
\and
G.~Qu\'em\'ener\inst{1}\fnmsep\thanks{\email{quemener@lpccaen.in2p3.fr}}\fnmsep\href{https://orcid.org/0000-0001-6703-6655}{\includegraphics[height = 2.2ex]{fig/ORCIDiD_exponent.png}}
\and
P.~Delahaye\inst{2}\fnmsep\href{https://orcid.org/0000-0002-8851-7826}{\includegraphics[height = 2.2ex]{fig/ORCIDiD_exponent.png}}
\and
X.~Fl\'echard\inst{1}\fnmsep\href{https://orcid.org/0000-0003-3754-9083}{\includegraphics[height = 2.2ex]{fig/ORCIDiD_exponent.png}}
\and
E.~Li\'enard\inst{1}\fnmsep\href{https://orcid.org/0000-0003-3748-4021}{\includegraphics[height = 2.2ex]{fig/ORCIDiD_exponent.png}}
\and
B.M.~Retailleau\inst{2}
}

\institute{
Normandie Univ, ENSICAEN, UNICAEN, CNRS/IN2P3, LPC Caen, 14000 Caen, France 
\and
GANIL, CEA/DSM-CNRS/IN2P3, Bd Henri Becquerel, 14000 Caen, France
}

\date{Received: date / Revised version: date} 
\date{\today}

\abstract{
In the frame of the project MORA (Matter's Origin from the Radio Activity of trapped and oriented ions), 
a transparent axially symmetric radio-frequency ion trap (\morat) was designed in order to measure the 
triple correlation parameter $D$ in nuclear $\beta-$decay of laser-polarised ions. 
The trap design was inspired from the \lpct geometry, operated at GANIL from 2005 to 2013. 
In a real (non-ideal) Paul trap, the quadrupole electric potential is not 
perfect leading to instabilities in ion motion and therefore affecting the 
overall trapping efficiency. 
This paper presents a numerical method aiming to optimise the geometry of a trap. It is applied to \morat in order to improve the trapping efficiency and to enlarge the axial transparent solid angle compared to \lpct.
In the whole optimisation process, numerical computation of electric potential and field was carried out using an electrostatic solver based on boundary element method (BEM). The optimisation consisted in minimising an objective function (fitness function) depending on higher order multipoles of the potential. 
Finally, systematic changes of trap dimensions and electrode displacements were applied to investigate geometrical effects on the potential quality.
\PACS{ { }{Paul trap, Spherical harmonics, Laplace's solver, Boundary element method} }
} 

\authorrunning{M.~Benali {\it et al.}}
\titlerunning{Geometry optimisation of a transparent axisymmetric ion trap for the MORA project}

\maketitle

%
%
\section{Introduction}
\label{sec:Introduction} 
  
Precision measurements in nuclear $\beta-$decay provide a remarkable tool to improve the accuracy of 
Standard Model (SM) parameters and to search for New Physics (NP) beyond, at the low energy frontier
\cite{Gonzales:2019}. 
In particular, the search for new sources of CP (Charge Parity) violation
is one of the requirements to explain the matter$-$an\-ti\-mat\-ter asymmetry observed in the universe, 
according to Sakharov's criteria~\cite{Sakharov:1967}. 
This search can be achieved in nuclear $\beta-$decay by measuring the triple correlation between the 
parent nucleus spin ($\vec{J}$), electron momentum ($\vec{p}_e$), and neutrino momentum ($\vec{p}_{\nu}$): 
$D \left\langle \vec{J}\right\rangle \cdot(\vec{p}_e\times \vec{p}_{\nu})$. 
This triple correlation is sensitive to T (Time) reversal violation and thus to CP violation
thanks to CPT conservation. Such a violation would be quantified by a non-zero $D$ value, experimentally
determined from an asymmetry in the  $\beta$-recoil angular distribution measured in a plane
perpendicular to $\vec{J}$ for two opposite directions of this nucleus orientation.
In this context, the new project MORA~\cite{Delahaye:2018} aims to measure the triple-correlation $D$ coefficient in the 
$\beta$-decay of laser-polarised $^{23}\mbox{Mg}^+$ and $^{39}\mbox{Ca}^+$ ions confined in a Paul 
trap applying a quadrupole radio frequency (RF) field. The use of Paul traps
is considered as an innovative technique in precision  measurements  of correlation coefficients 
in nuclear $\beta-$decay in the SM framework~\cite{Sternberg:2015, Brodeur:2016}. One example is the transparent Paul trap \lpct used 
in the measurement of the $\beta-\nu$ correlation coefficient,  $a_{\beta\nu}$, in the decay of 
different nuclei~\cite{Ban:2013, Fabian:2014, Lienard:2015, Delahaye:2019}.

\begin{figure}[hbt!]
\centering
\includegraphics[width=0.8\linewidth]{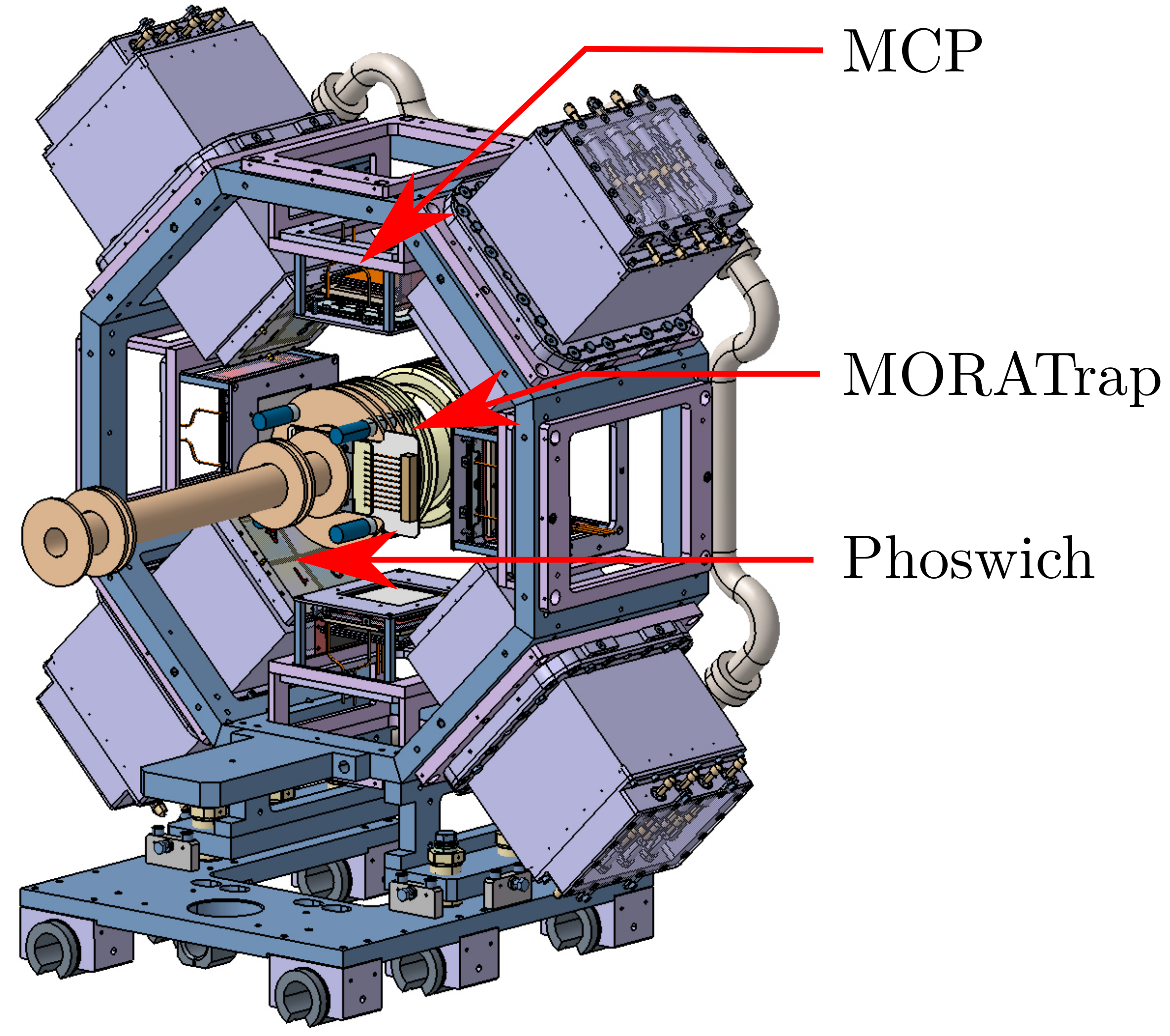}
\caption{\morat setup overview showing the location of the ion trap in the centre surrounded by 4 beta detectors 
(Phoswich) and 4 recoil ion detectors (MCP).}
\label{fig:setup}
\end{figure}

The central element of the MORA apparatus is a transparent Paul trap (\morat), which will be used to confine 
singly charged radioactive ions, coupled to a laser system allowing to polarise 
the nucleus by optical pumping. 
As shown in Fig.~\ref{fig:setup}, \morat, which will be installed in a vacuum
 chamber, is surrounded by 
four pairs of electron and recoil ion detectors arranged alternately in an 
octagonal geometry in the 
azimuthal plane of the trap and allowing close to $2\pi$ azimuthal coverage. 
Two annular silicon detectors (not visible on the figure) located on the trap 
axis will monitor the polarisation degree thanks to $\beta$ asymmetry measurement.

The RF potential generated in the trapping volume or region of interest (ROI) is
not perfectly a quadrupole but contains some small amplitude, higher order electric multipole components which disturb the ion's motion. 
Since the potential in the ROI depends on the electrodes shape and on the applied voltages, an optimisation of the trap geometry is mandatory to reduce the higher order harmonics and to generate an optimised quadrupole potential.  

In this work, we will describe an efficient method to optimise the \morat 
geometry, using two electrostatic solvers 
developed at LPC Caen. Our trap is a three dimensional transparent Paul trap,
 inspired from the \lpct geometry, 
where the electrodes have been modified from an ideal Paul trap to allow the detection of $\beta-$ decay products in a large solid angle around the azimuthal plane
and obtain an efficient injection of the ion bun\-ches. By optimising the geometry, we aim to reach a quadrupole
potential of higher quality, to minimise the ion losses from the trap and to increase the trapping lifetime and the space charge capacity. Such improvements are mandatory to reach the statistics required in the MORA experiment 
to search for NP. 

The potential of an ideal Paul trap is commonly defined in cylindrical ($r$, $z$) coordinates as:
\begin{equation}
V_{\mathrm{ideal}}(r,z,t) \, = \, V_0 \, \cos(\Omega t)\, \frac{r^2 \, - \, 2 \, z^2}{2 \, r_0^2},
\label{eq:Videal}
\end{equation}
where $V_0$ is the amplitude of the RF voltage, $\Omega = 2 \pi f$ its pulsation, and $r_0$ a 
distance parameter. 3D Paul traps usually approximate the ideal Paul trap by using electrodes 
consisting of one ring and two end caps, whose surfaces form truncated hyperboloids (see for 
example Fig.~1 in \cite{Delahaye:2019}). In such a configuration, $r_0$ is the distance of the 
ring surface to the centre of the trap. In geometries which significantly depart from the ideal 
trap, one can define an effective trapping radius $r_{\mathrm{eff}}$ which contains the region 
where the quadrupole field still sufficiently dominates, so that ion trajectories are still 
stable in standard conditions defined for the ideal potential. This region is defined in 
\cite{Delahaye:2019} as the region for which the contribution to the potential from harmonics 
of higher order than the quadrupole one, are still below a few percents.
The conditions for stability of an ion of mass $m$ and charge $q$ in an ideal Paul trap, in 
the absence of DC field, are defined with respect to the Mathieu parameter $q_z$:
\begin{equation}
q_z \, = \, \frac{4 \, q \, V_0}{m \, r_0^2 \,\Omega^2}.
\label{eq:MatPar}
\end{equation}
In the first stability region, $|q_z|<0.908$ \cite{March:2005}. In the pseudo-potential approximation limit, valid for small $q_z$ values, one can define a pseudo-potential depth $D_z$ (resp. $D_r$) for 
the $z$ (resp. $r$) dimension:
\begin{equation}
D_z \, = \, \frac{q_z \, V_0}{8} \, = \,2 \, D_r.
\label{eq:PseuPot}
\end{equation}  
In this model, the maximal charge density $\rho_{\mathrm{max}}$ the Paul trap can hold is
\begin{equation}
\rho_{\mathrm{max}} \, = \, \frac{3 \, \varepsilon_0 \, D_z}{z_0^2},
\label{eq:rhomax}
\end{equation} 
where $\varepsilon_0$  is the vacuum permittivity and $z_0 \, = \, r_0 / \sqrt{2}$ is the distance of the end caps from the trap centre.
From Eqs.~(\ref{eq:MatPar}) to (\ref{eq:rhomax}) and considering the fact that experimentally one usually fixes a Mathieu parameter in the middle of the stability diagram $q_z \,\simeq \, 0.4$~\cite{Delahaye:2019}, the maximal charge capacity $Q_{\mathrm{max}}$ a trap can hold directly relates to the product $r_0 V_0$ via the formula:
\begin{equation}
Q_{\mathrm{max}} \, = \, \frac{1}{2}  \pi  \varepsilon_0  q_z  r_0  V_0
\label{eq:Qmax0}
\end{equation}
in the case of an ideal trap, and  
\begin{equation}
Q_{\mathrm{max}}=\frac{1}{2}  \pi \varepsilon_0 q_z r_{\mathrm{eff}} V_{\mathrm{eff}}
\label{eq:Qmax}
\end{equation}
in the case of a Paul trap with a limited quadrupole region of radius $r_{\mathrm{eff}}$, where $V_{\mathrm{eff}} \, = \, V_0 \frac{r_{\mathrm{eff}}^2}{r_0^2}$ is the maximum potential at the radius $r_{\mathrm{eff}}$. In order to optimise the charge capacity of the trap for the MORA experiment, the optimisation procedure presented in this article therefore aims at enlarging the trapping region, of radius $r_{\mathrm{eff}}$, given some space constraints on the size of the trap.

Sec.~\ref{sec:LaplaceSolver} presents the electrostatic solvers necessary to compute the potential from an 
electrode set and Sec.~\ref{sec:LaplaceSeries} introduces the harmonics series expansion describing 
this potential, solution of Laplace's equation. Sec.~\ref{sec:ObjectiveFunction} details the objective function 
used in the optimisation procedure whose results are presented and commented in Secs.~\ref{sec:OptimisationResults} 
and \ref{sec:Comparison}.
Finally, the effects of electrode misalignment on the potential quality are 
investigated in Sec.~\ref{sec:DesignSensitivity}.

%
%
\section{Laplace's solver}
\label{sec:LaplaceSolver}

The shape of the electric potential generated within a trap depends on both the electrode 
geometry and their corresponding applied voltages. In order to estimate and optimise this potential, 
it needs to be accurately computed within the ROI (the trapping region). This is achieved by solving Laplace's equation in this ROI.
For that purpose, the most common approaches use either the finite element method (FEM) or the finite 
difference method (FDM) which require to mesh both the electrodes and the free space volume. For our study, 
we have used a home made C\texttt{++} Laplace's solver based on the boundary element method (BEM). 
This solver has been developed by one of us and thoroughly validated with analytic examples and 
expensive commercial software like {\textsc{simion}} (FDM) or {\textsc{comsol}} (FEM).
Compare to FEM and FDM, BEM exhibits several advantages: it only requires to mesh the surface of the
electrodes and is thus able to achieve higher precision with less computation time and memory requirements. 
This is especially true for geometries involving electrodes with large aspect ratios ({\it e.g.} a very thin thickness for very large length and width) for which
the 3D meshing quality would require special care. In addition, BEM can easily deal with open systems as 
the boundary conditions are inherent to the formalism.

Even if our BEM program also handles dielectric materials, in order to simplify its description, here we shall 
only concentrate on sets of electrodes with applied voltages. A setup is described by $N_e$ electrodes each represented by its surface. The surface of 
electrode $e \,(e \in [1, N_e])$ is meshed in $n_{e}$ flat polygonal cells (triangles, quadrangles, ...).
The full setup is therefore represented by the set of cells $\mathcal{C} = \left\{ \mathcal{C}_i: i \in [1, N] \right\}$ where
$N = \sum_{e=1}^{N_e} n_{e}$ is the total number of cells,
each with an {\it a priori} unknown associated surface charge density $\sigma_i$ assumed constant over the 
cell surface. The $n_e$ cells of electrode $e$ obviously share the same potential applied on the whole electrode. 
The potential $V_i$ of cell $\mathcal{C}_i$ centred at position $\mathbf{r}_i$ is related to the 
surface charge densities $\sigma_j$ of cells $\mathcal{C}_j$ ($j \in [1, N]$) through the superposition 
principle as:
\begin{equation}
  V_i \, = \, \sum_{j=0}^{N} \, \frac{\sigma_j}{4\pi\varepsilon_0} \, \int_{\mathcal{C}_j} \frac{1}{\| \mathbf{r}_i - \mathbf{r'} \|} \mathrm{d}^2\mathbf{r'} \, = \, \sum_{j=0}^{N} \, Q_{ij} \, \sigma_j,\label{eq:BEM}
\end{equation}
where the integral over the surface of cell $\mathcal{C}_j$ only depends on $\mathcal{C}_j$ shape and on 
the relative position of the target point $\mathbf{r}_i$  with respect to $\mathcal{C}_j$ location,
as $\sigma_j$ is assumed constant over the whole surface of cell 
$\mathcal{C}_j$. An analytic formula has been derived for the integral in Eq.~(\ref{eq:BEM}). 
This formula is too complicated to be discussed here and shall be published in a separate article, 
but it should be emphasised that special care has been taken to 
suppress numerical instabilities in its evaluation especially when $i = j$ where numerical divergence may 
occur.  Eq.~(\ref{eq:BEM}) can be written for each cell 
$\mathcal{C}_i$, leading to the following set of $N$ equations:
\begin{equation}
   \begin{pmatrix} 
    V_1\\
    V_2\\
    \vdots\\
    V_N
    \end{pmatrix}
    =
    \begin{pmatrix}
    Q_{1,1} & Q_{1,2}& \cdots  & Q_{1,N}\\
    Q_{2,1} & Q_{2,2}& \cdots  & Q_{2,N}\\
    \vdots & \vdots & \ddots &\vdots \\ 
    Q_{N,1} & Q_{N,2}& \cdots  & Q_{N,N}
    \end{pmatrix}
    \begin{pmatrix}
    \sigma_1\\
    \sigma_2\\
   \vdots\\
    \sigma_N
    \end{pmatrix},
    \label{eq:matrixBEM}
\end{equation}
where $V_i$ are the known potentials applied on electrodes and $\sigma_i, i \in [1, N]$ are the unknown surface 
charge densities.
The matrix elements $Q_{ij}$ are computed for a given electrode assembly/geometry leading to a dense square 
matrix on the contrary to FEM and FDM which deal with sparse matrices of much larger dimensions. The presence 
of a dense matrix requires special algorithms to solve Eq.~(\ref{eq:matrixBEM}) for the $\sigma_i$: a direct 
solver such as LU decomposition can be used to obtain an exact solution or an approximated solution can also 
be computed more rapidly with iterative solvers such as CMRH~\cite{Sadok:1999}. Once the charge densities 
have been determined, the potential and field
components can be evaluated at any location $\mathbf{r}$ in space without interpolation contrary to the FEM and FDM
approaches:
\begin{eqnarray}
  V(\mathbf{r}) & = & \sum_{i=0}^{N} \, \frac{\sigma_i}{4\pi\varepsilon_0} \, \int_{\mathcal{C}_i} \frac{1}{\| \mathbf{r} - \mathbf{r'} \|} \mathrm{d}^2\mathbf{r'}  \label{eq:BEM2} \\
  \mathbf{E}(\mathbf{r}) & = & \sum_{i=0}^{N} \, \frac{\sigma_i}{4\pi\varepsilon_0} \, \int_{\mathcal{C}_i} \frac{\mathbf{r} - \mathbf{r'}}{\| \mathbf{r} - \mathbf{r'} \|^3} \mathrm{d}^2\mathbf{r'} \label{eq:BEM3}
\end{eqnarray}
As for Eq.~(\ref{eq:BEM}), integrals in Eqs.~(\ref{eq:BEM2}) and (\ref{eq:BEM3}) are analytically evaluated over the surface of each cell.

\begin{figure}[hbt!]
\centering
\includegraphics[width=0.95\linewidth]{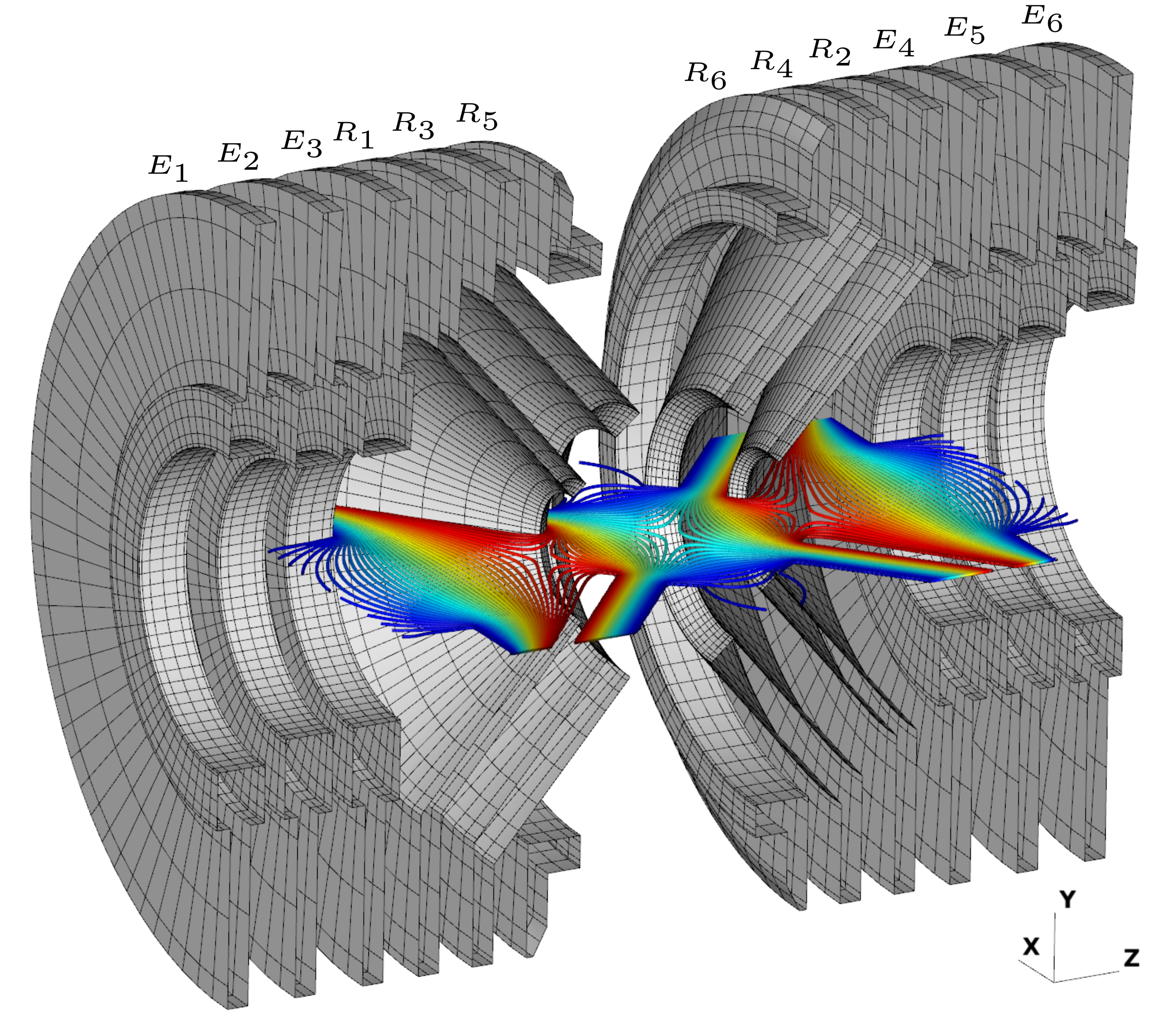}
\caption{{\textsc{gmsh}} cross section view of \morat showing the electrodes along with iso-potential lines obtained with
{\textsc{electrobem}}. The trap consists of three ring electrode pairs ($R_1-R_2$, $R_3-R_4$ and $R_5-R_6$) surrounded by two Einzel lens triplets ($E_1-E_3$ and $E_4-E_6$). A more detailed description is given in Sec.~\ref{sec:OptimisationResults}. 
The thin black lines on the electrodes delimit the perimeter of the polygonal cells 
used to solve the electrostatic problem.}
\label{fig:trap_and_potential}
\end{figure}

\begin{figure}[hbt!]
\centering
\includegraphics[width=0.8\linewidth]{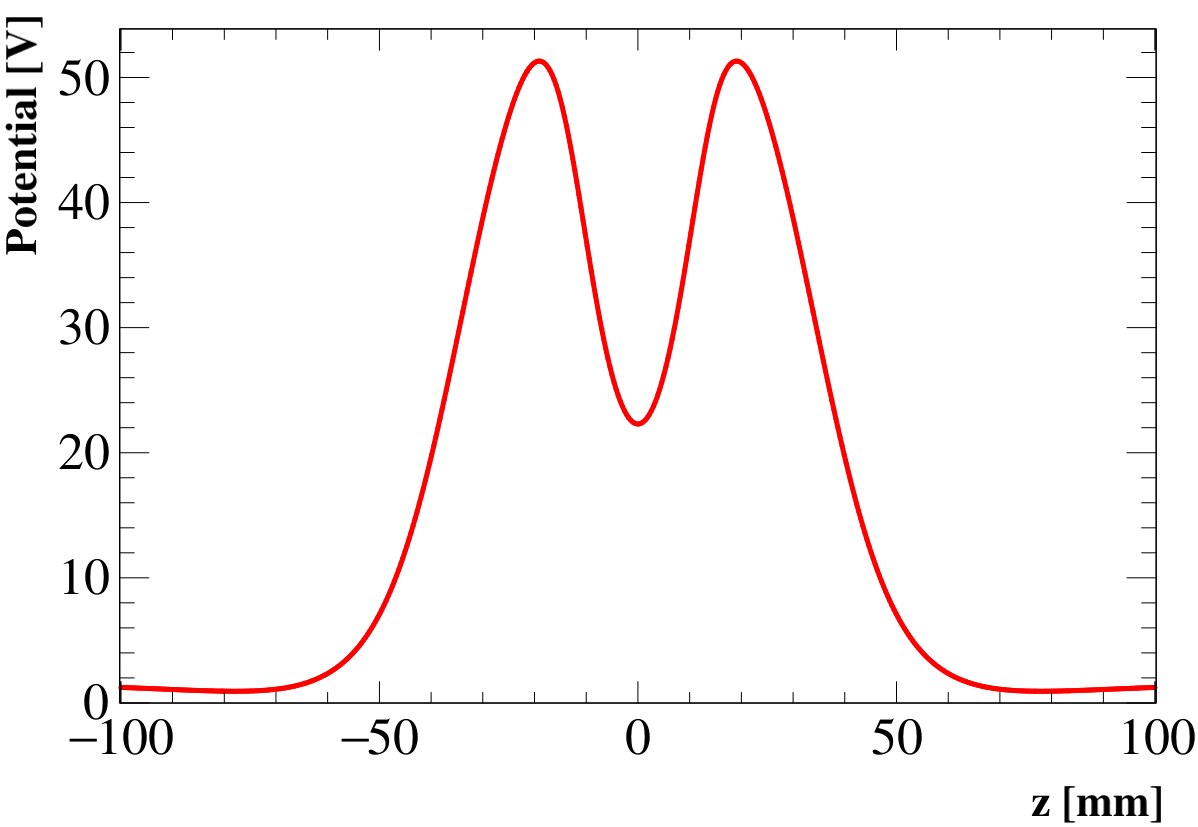}
\caption{{\textsc{root}} view of the potential along \morat axis ($r = 0$) computed in {\textsc{axielectrobem}} for 60~V applied on inner electrodes $R_1$ and $R_2$, all other ones being grounded.}
\label{fig:potential_alongz}
\end{figure}

For our studies we have used two versions of the solver: a full 3D solver, called {\textsc{electrobem}}, and a derived 
version, {\textsc{axielectrobem}} for axially symmetric problems\footnote{In axisymmetric problems, the matrix elements 
$Q_{ij}$ and other surface integrals are computed differently making use of complete elliptic integrals.}. In {\textsc{electrobem}}, a complex 
3D setup can be modelled with a thorough set of geometric functions handling predefined shapes (flat polygons, ring, 
cylinder, cone, torus, ...), rotations, translations and symmetry planes as well as fine tuning of mesh properties. 
This last point is crucial for the precision of the solution. The software can also import pre-computed meshes from 
the open source, reliable and user friendly {\textsc{gmsh}} software~\cite{Geuzaine:2009}. Both {\textsc{gmsh}} and 
{\textsc{root}}~\cite{Root:1997} may be used to display the geometry. Electric potentials are applied to electrodes and then, once the electrostatic problem has been solved, 
a complete set of plotting/mapping functions allows to display and/or export the potential 
and field components in 1D, 2D and 3D as illustrated for instance in Figs.~\ref{fig:trap_and_potential} and \ref{fig:potential_alongz}. In {\textsc{axielectrobem}}, any shape axially symmetric around $z$-axis 
can be simulated by predefined shapes or by functions ($r = f(z)$). The meshing is done along longitudinal $z-$axis 
and radial $r-$axis. In this version, the number of cells necessary to model a given axisymmetric setup is 
substantially smaller than it would be in the 3D version for the same geometry, resulting in a much smaller 
computation time. {\textsc{axielectrobem}} is thus particularly well suited for the optimisation of the axially 
symmetric \morat and was used in a first phase of our study, whereas the 3D version served in a second phase 
to estimate the effects, on the trapping potential quality, of possible mechanical misalignment or machining 
precision which break the setup axisymmetry. Depending on the version, 2D or 3D, the potential inside the trap 
is expanded differently in terms of multipole coefficients, as shown  in next section and in appendix \ref{app:HomogPoly}.


%
%
\section{Laplace's equation: spherical harmonics series expansion}
\label{sec:LaplaceSeries}

In a source-free region, {\it i.e.} in the absence of charges, the electric potential satisfies Laplace's equation
$\Delta V(\rho, \theta, \varphi) \, = \, 0$ with $(\rho, \theta, \varphi)$ being the usual spherical coordinates. 
The solution of this equation~\cite{edurand}, may be expressed in terms of a uniformly convergent Laplace
series also known as spherical harmonics series expansion in the following way:
 \begin{eqnarray}
 V(\rho,\theta,\varphi) & = & \sum_{n=0}^{\infty}\sum_{m=0}^{n} \! \sqrt{\frac{2n+1}{4\pi}\frac{(n-m)!}{(n+m)!}} \left(\frac{\rho}{R_0}\right)^{\!\!n} \!\! P^m_n(\cos\theta)\nonumber \\
 & \times & \left[\alpha_{nm} \cos(m\varphi) - \beta_{nm} \sin(m\varphi)\right] \nonumber \\
& = & \sum_{n=0}^{\infty}\sum_{m=0}^{n} \left(\frac{\rho}{R_0}\right)^{\!\!n} \!\! P^m_n(\cos\theta)\nonumber \\
 & \times & \left[A_{nm} \cos(m\varphi) - B_{nm} \sin(m\varphi)\right],
 \label{eq:SH}
 \end{eqnarray}
where $P^m_n(\cos\theta)$ are the associated Legendre functions of the first kind, of degree $n$ and order $m$, and 
$A_{nm}$ (resp. $B_{nm}$) are the normal (resp. skew) spherical harmonics coefficients ($\alpha_{nm}$ and $\beta_{nm}$ being the corresponding unnormalised harmonics). $R_0$ 
is the convergence radius of the above series expansion, it must be smaller than the distance between the origin 
($\rho = 0$) and the closest electrode: the region of convergence or the ROI must remain a source free 
region.

In axisymmetric cases with azimuthal symmetry, the potential does not depend on the $\varphi$ angle and therefore 
we only keep terms with $m=0$ in the above expansion. Dropping the $m$ index and the $\varphi$ dependence, 
Eq.~(\ref{eq:SH}) reduces to:
\begin{eqnarray}
  V(\rho,\theta) & = & \sum_{n=0}^{\infty} \alpha_{n} \sqrt{\frac{2n+1}{4\pi}} \left(\frac{\rho}{R_0}\right)^{\!\!n} P_n(\cos\theta) \nonumber \\
  & = & \sum_{n=0}^{\infty} A_{n} \left(\frac{\rho}{R_0}\right)^{\!\!n} P_n(\cos\theta) 
  \label{eq:CH}
\end{eqnarray}
with $P_n(\cos\theta)$ the Legendre polynomial of degree $n$ and $A_n$ the harmonic coefficient of order 
$n$ ($\alpha_{n}$ is the corresponding unnormalised harmonics). $A_0$ is a constant term for the potential and therefore it does not appear in the field components 
which enter the ion equations of motion. $A_1$ is called the dipole harmonics, $A_2$ the quadrupole 
one and so on for higher orders. The harmonics $A_{2k+1}$ with $k=0, 1, 2, \dots$ are said to belong to the
{\it dipole series} whereas harmonics $A_{4k+2}$ with $k=0, 1, 2, \dots$ are said to belong to the 
{\it quadrupole series}. 
Because of symmetry reasons, a potential exhibiting a quadrupole behaviour, is more likely to have harmonics 
from the quadrupole series ($A_2, \, A_6, \, A_{10}, \dots$) dominant over other higher order terms. One 
should also notice that the general trend of $A_n$ is to decrease with increasing $n$ and that the contribution 
from $A_n$ to the potential is weak close to the origin and increases as $\rho^n$. 
If, in addition to the axial symmetry, a setup exhibits a symmetry plane at $z = 0$, then only harmonics 
$A_n$ with even order $n$ will be present in the expansion. The harmonic coefficient $A_n$
is obtained by integration over the sphere $S$ of radius $R_0$ by:
\begin{equation}
   A_n \, = \, \frac{2 n + 1}{4 \pi} \!\!
\int_{0}^{2\pi} \!\!\mathrm{d}{\varphi} \int_{0}^{\pi} \!\! P_n(\cos\theta) \,\, V(R_0, \,\theta) \, \sin\theta \, \mathrm{d}{\theta}.
\end{equation}
After integration over $\varphi$, and assuming a symmetry plane at $z = 0$, $A_n$ becomes:
\begin{equation}
   A_n \, = \, (2 n + 1) \, \int_{0}^{\frac{\pi}{2}} \! P_n(\cos\theta) \: V(R_0, \,\theta) \, \sin\theta \, \mathrm{d}{\theta}. 
   \label{eq:An}
\end{equation}
In order to determine the ${A}_n$ coefficients for a given electrode configuration, Eq.~(\ref{eq:An}) 
is numerically integrated using a 64-nodes Gauss-Legendre quadrature with the potential 
$V(R_0, \,\theta)$ calculated on the circle of radius $R_0$ at the Gauss-Legendre nodes ($\theta_i$) using the BEM solver described in previous 
section. The extracted harmonic spectrum $\{A_n\}$ will then be  used in the minimisation procedure of an objective function to optimise the trap geometry.
A similar approach, combined with FFT (Fast Fourier Transform), is used in 3D to determine the harmonics $A_{nm}$ and 
$B_{nm}$, but a sphere of radius $R_0$ is used instead of a circle to sample the potential $V(R_0, \theta, \varphi)$. 
The potential is determined on a $\theta-\varphi$ grid of $180 \times 360 = 64800$ points.

Let us now introduce the objective function used in this study.

%
%
\section{Objective function}
\label{sec:ObjectiveFunction}

\begin{figure}[hbt!]
 \centering
 \includegraphics[width=0.95\linewidth]{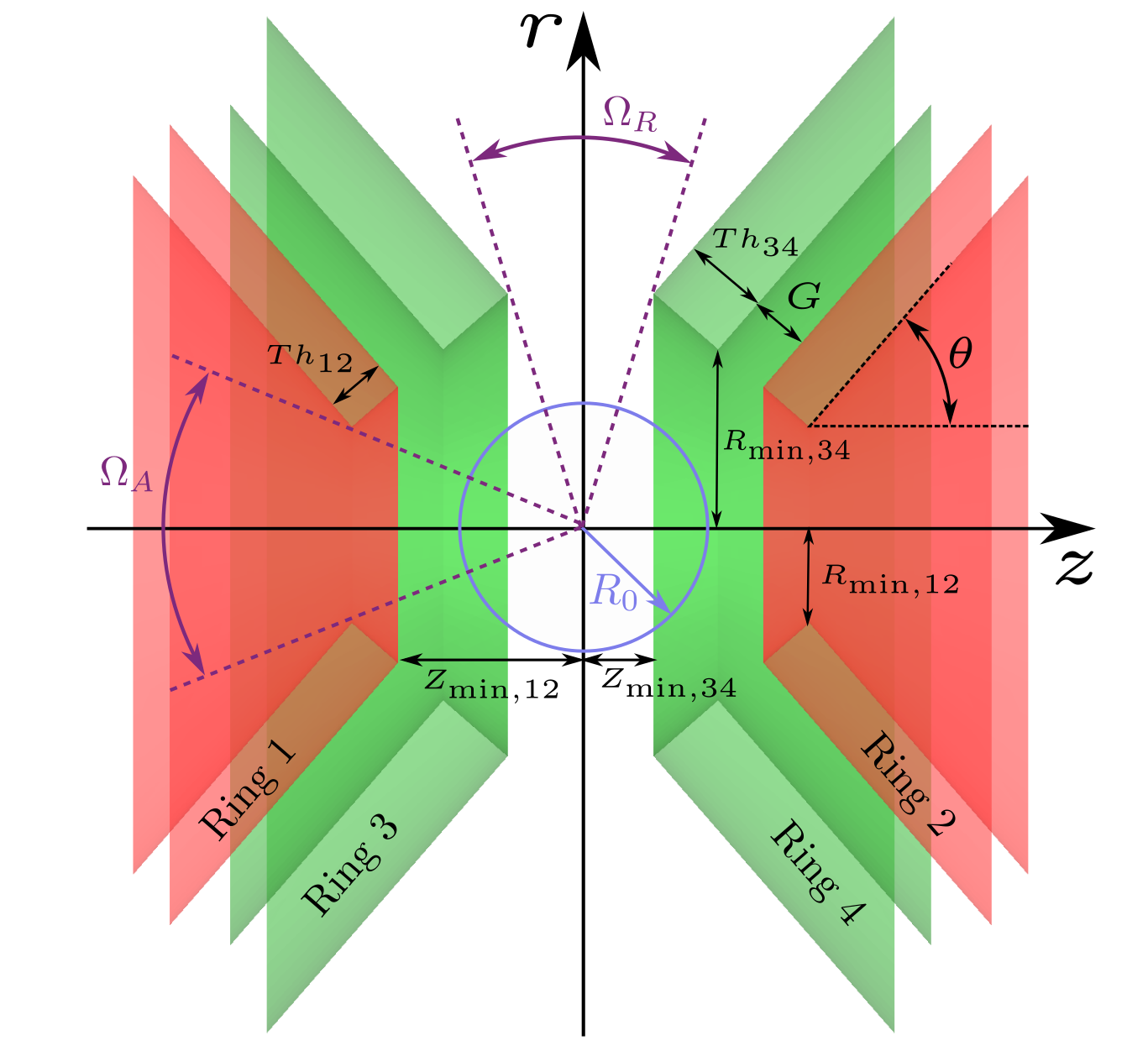}
 \caption{ Definition of the geometric parameters used to optimise the trap: 
 axial distance $Z_{min}$ from the radial axis to the closest point on a ring
 electrode, minimal radial distance $R_{min}$, radial thickness of the electrode 
 $Th$, gap $G$ between two successive electrodes and cone angle $\theta$ from 
 $z$-axis. $\Omega_A$ and $\Omega_R$ are respectively the axial and 
 radial angular openings. The circle of radius $R_0$ used to compute the spherical harmonics is   
 also drawn.}
 \label{fig:FreeParameters}
 \end{figure}

As already mentioned, an ideal Paul trap requires a pure quadrupole potential, {\it i.e.} $A_2$ as large as 
possible while $A_n = 0 \,\,\, \forall n > 2$. In practice, this is almost impossible to achieve and higher 
order harmonics are present in the potential. The present study aims at 
optimising the geometry of the \morat electrodes
by minimising higher order harmonics in order to avoid the loss of
trapped ions because of trajectory instabilities induced by these higher order terms. As shown in 
\cite{Delahaye:2019}, a necessary condition to 
suppress ions losses is to keep the relative contribution to the potential from
higher harmonics to less than a few \%, say $2\%$, in the ROI (trapping region). Neglecting the constant term $A_0$ which does not contribute to the field and only considering 
$n$-even terms because of planar symmetry at $z = 0$, this condition translates into a relative difference to 
the quadrupole $A_2$ as:
\begin{equation}
 \frac{ \sum_{n=2}^{n_{\mathrm{max}}}\!\left(\!\frac{\rho}{R_0}\!\right)^{\!n} \!\! \left|A_{n}\right| \!-\! \left(\!\frac{\rho}{R_0}\!\right)^{\!2}\left|A_{2}\right|}{\left(\!\frac{\rho}{R_0}\!\right)^{\!2} \!\!\left|A_{2}\right|} \!=\!\sum_{n>2}^{n_{\mathrm{max}}}\!\left(\!\frac{\rho}{R_0}\!\right)^{\!\!n-2} \!\! \left|\frac{A_{n}}{A_2}\right| \leqslant \, 0.02,
 \label{eq:2pc}
\end{equation}
where the series expansion is truncated to $n_{\mathrm{max}}$ 
whose value will be discussed in next section.
Choosing $2\%$ as an upper limit is therefore 
equivalent to find a root $\rho_{2\%}$ of a polynomial of degree $n_{\mathrm{max}}-2$ in $\rho$. 
Optimising the trap performances amounts to maximise both the $A_2$ term contribution as well
as $\rho_{2\%}$. Both operations can be achieved by minimising the following 
objective or fitness function:

\begin{equation}
  f(\mathbf{a}) \, = \, \frac{1}{\left(\frac{\rho_{2\%}(\mathbf{a})}{R_0}\right)^{\!\!2} A_2(\mathbf{a})} ,
\label{eq4}
\end{equation}
where $\rho_{2\%}(\mathbf{a})$ and $A_2(\mathbf{a})$ explicitly show their dependence on the geometry via the vector $\mathbf{a}$ which contains the 
different free parameters describing the electrodes. To keep the  shape of the electrodes
as simple as possible and therefore to facilitate their machining at LPC Caen, we have chosen the following parameters for some conical electrodes as illustrated in Fig.~\ref{fig:FreeParameters}:

\begin{itemize}
\item $Z_{min}$, the minimal axial distance from the trap $r$-axis to the electrode.
\item $R_{min}$, the innermost radius which is the minimal radial distance from the trap $z$-axis to the electrode.
\item the cone angle $\theta$.
\item the electrode section thickness $Th$.
\end{itemize}

Some other parameters can be derived from these quantities like the gap $G$ between two electrodes,
their outermost radii as well as the radial and axial angular acceptances respectively $\Omega_R$ 
and $\Omega_A$. These could help imposing some mechanical constrains in the minimisation procedure.
 
For some exploratory work, we had envisaged and test\-ed more complicated shapes for the electrodes, including {\it e.g.} some spli\-nes to describe parts of their section, but the whole study show\-ed that simpler shapes could suit our precision needs as we shall see in next section. 

%
%
\section{Optimisation process and results}
\label{sec:OptimisationResults}

Maximising the measured statistics with \morat requires not only the trapping of as many ions
as possible during a period as long as possible, but also a large angular aperture $\Omega_R$
in the radial direction where the $\beta$ and recoil ions detectors are located. In addition, 
the $D$ measurement relies on the knowledge of the ion cloud polarisation which shall be continuously 
monitored during the experiment~\cite{Delahaye:2018}.
This will be performed with two annular silicon detectors located along the trap $z$-axis before 
and after the electrodes. The central hole in these Si detectors 
lets the ion beam entering and exiting the trap. The axial angular acceptance of these
detectors directly impacts the precision on the polarisation measurement and is constrained by 
the trap angular opening $\Omega_A$. All the above constrains have been taken care of in the 
optimisation process.
\begin{figure}[hbt!]
 \centering
 \includegraphics[width=0.65\linewidth]{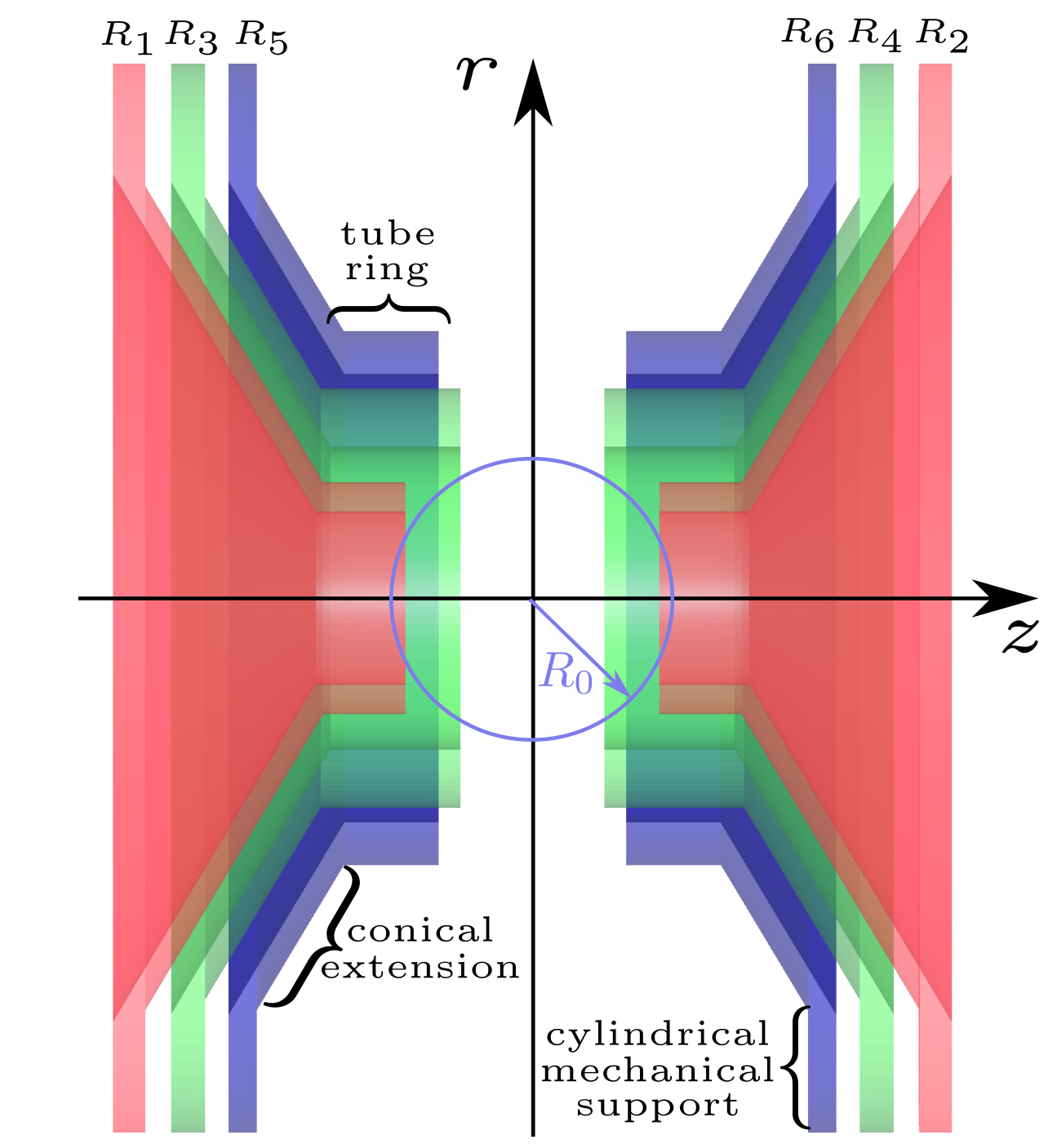}
 \caption{Cross section view of \lpct geometry. There are three pairs of ring/tube electrodes 
 with conical  extensions and cylindrical mechanical support. $R_1$ and $R_2$ 
 are the RF electrodes, $R_3$ and $R_4$ are used to inject and extract the ions and the external
 electrodes $R_5$ and $R_6$ are used to correct the potential generated along the $r$-axis and 
 to steer the trajectories of the decay products. The circle of radius $R_0 = 10$~mm gives
 the overall scale. }
 \label{fig:xsv_LPCTrap}
\end{figure}   

\begin{figure}[hbt!]
  \centering
  \includegraphics[width=0.95\linewidth]{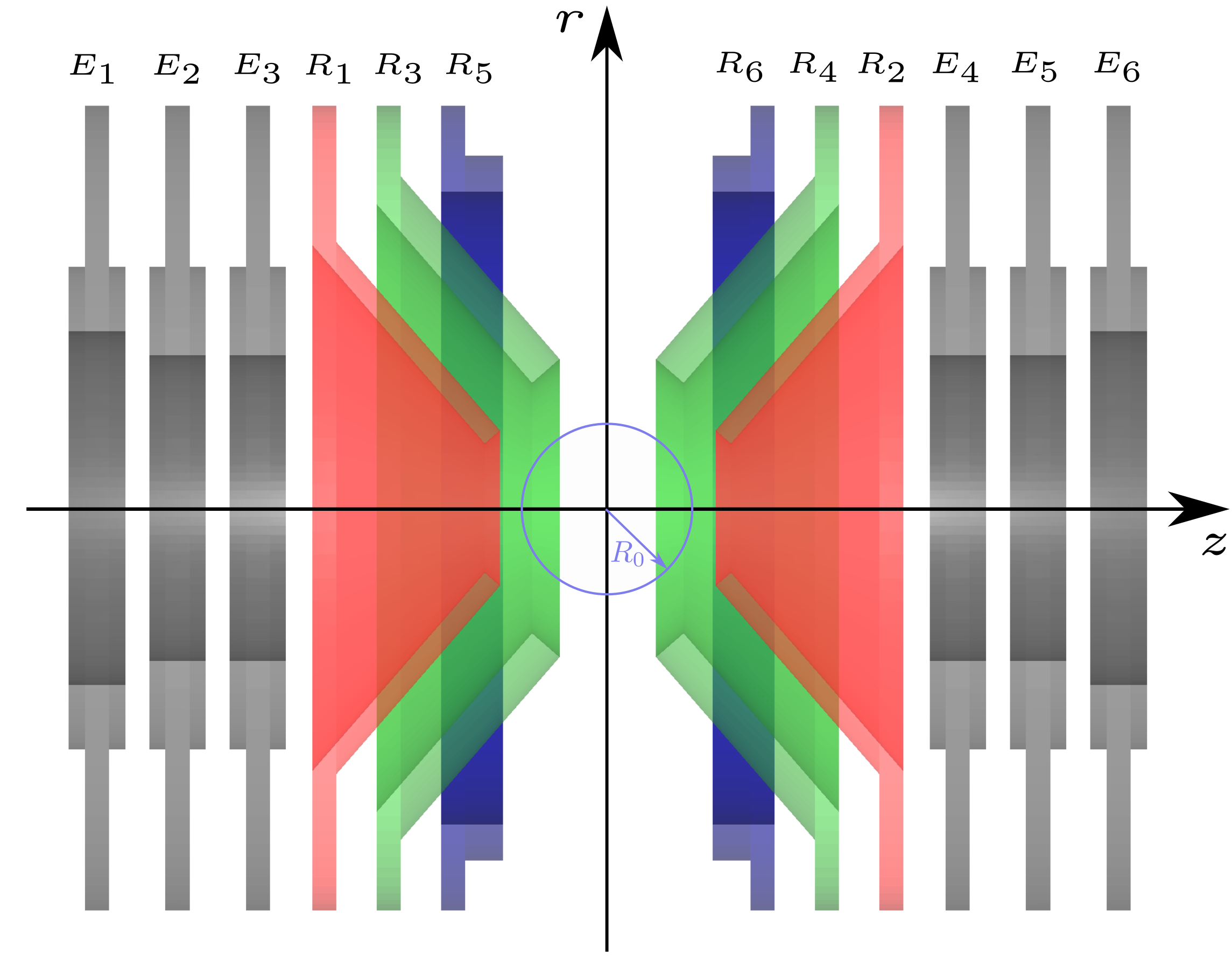} 
  \caption{Cross section view of the optimised \morat geometry. A RF 
  voltage is applied to the innermost ring pair ($R_1$, $R_2$), while the middle and outer electrodes, 
  as well as the Einzel lenses  are grounded. $z$-axis is along the beam injection and extraction. 
  The geometry is symmetric with respect to the $x-y$ plane and is invariant around the $z$ axis.
  The circle of radius $R_0 = 10$~mm gives the overall scale, but more precise dimensions are given in Table~\ref{tab:properties}. }
    \label{fig:xsv_MORATrap}
\end{figure}

A simplified version of \lpct (Fig.~\ref{fig:xsv_LPCTrap}) was used as a starting point to look for 
an optimal geometry\footnote{In reality some parts of \lpct electrodes are not axisymmetric at large 
distance from the ROI, but we simplified their geometry and made them cylindrical in the simulation.} 
as it already provides a large enough radial acceptance 
$\Omega_R$~\cite{Fabian:2014}. This starting geometry respects axial symmetry and 
also possesses a planar symmetry at $z = 0$. 
However, on the contrary to \lpct which uses three pairs of tube or cylinder shaped electrodes, in order to 
satisfy \morat's requirement of a larger axial acceptance $\Omega_A$, conical shapes were 
chosen for the two innermost pairs of electrodes. A third pair of electrodes ($R_5-R_6$) was
added to screen the axially symmetric trap region from the octagonal set of detectors and 
associated collimators which might disturb the potential in the ROI (Fig.~\ref{fig:setup}). This outermost set of electrodes 
is tube shaped, as in the \lpct geometry. 
Two Einzel lens triplets, used to focus the 
incoming beam and to clean the trap after a measurement cycle, were also included in 
the simulation (Fig.~\ref{fig:xsv_MORATrap}). The shape of these lenses and of 
the outermost tube rings were optimised separately to fit some mechanical constrains and were kept fixed along 
the whole optimisation process of the two innermost electrode pairs. Because of their distance from the ROI 
they did not have any significant effect on the potential as will be seen in Sec.~\ref{sec:DesignSensitivity}.

To ease later comparisons with \lpct performances, it was decided to use, in the simulation, 
the same applied RF voltage of 60~V on the 
innermost pair of electrodes even if during \morat operation we shall use 100~V or more (all other
electrodes are grounded during the trapping period). 

To summarise, the parameters related to outermost electrodes and Einzel lens triplets were 
already fixed prior to the whole minimisation process and we have chosen to optimise seven free
 parameters linked to the geometry of the internal and middle 
electrodes (Fig.~\ref{fig:FreeParameters}):  the axial distances $Z_{min,12}$ and $Z_{min,34}$, the radial distances $R_{min,12}$ and $R_{min,34}$, 
a common cone angle $\theta$ and the thicknesses $Th_{12}$ and $Th_{34}$. The minimisation 
of the objective function $f(\mathbf{a})$ (Eq.~(\ref{eq4})) was performed within {\textsc{root-minuit}}~\cite{Root:1997, James:1994} 
and to further reduce the parameter space to be explored, some lower positive limits were determined by exploratory simulations
and set on the different distances $Z_{min}$ and $ R_{min}$.
The optimisation requires the computation of the potential harmonic spectrum.
To this purpose, the convergence radius
$R_0$  of the expansion was set to 10 mm which is about 5 times larger than the cloud radius
observed with \lpct. 
In addition, for the evaluation of $\rho_{2\%}$ from Eq.~(\ref{eq:2pc}), we have 
chosen to truncate the expansion at $n_{\mathrm{max}} = 18$, as the contribution
from the $A_{18}$ term, belonging to the quadrupole series, is already very small.
Indeed the absolute difference between the potential computed in 
{\textsc{electrobem}} and the potential synthesised from
harmonics up to order 18 is smaller than 2.5$\times 10^{-7}$~V
for the 64800 points used to compute the
 harmonics while the potential at these points varies from 15 to 37~V.

\begin{table}[hbt!]
\ra{1.2}
\begin{center}
\begin{tabular}{@{}lccc@{}}
\toprule
 &  & \bfseries{Rings}  &  \\
\cmidrule(lr){2-4}
\bfseries{Parameter}  & $\bm{R_1/R_2}$  & $\bm{R_3/R_4}$  & $\bm{R_5/R_6}$ \\
\midrule[0.25pt]
$\theta$ ($^{\circ}$)   &        49      &    49       &    0       \\
$Z_{\mbox{\scriptsize{min}}}$ (mm)         &      13.41     &     6.00    &   13.00    \\
Thickness (mm)          &       2.50     &     4.53    &    4.50    \\
Inner radius  (mm)      &       8.03     &    15.56    &   39.29    \\
Outer radius (mm)       &       9.67     &    18.53    &   43.79    \\
\bottomrule
\end{tabular}
\end{center}
\caption{Optimised \morat electrodes dimensions.}
\label{tab:properties}
\end{table}
After several iterations, {\textsc{minuit}} converged to a minimum of the  
objective function at $\rho_{2\%}=7.39$ mm.
The optimised geometric parameters of the electrodes corresponding  to this 
$\rho_{2\%}$ are presented in Table~\ref{tab:properties} and a
cross section view of \morat is shown in Fig.~\ref{fig:xsv_MORATrap}.
The precision on distances is limited to the machining tolerance (\SI{10}{\micro\metre}).
Effects of this limited precision are studied in Sec.~\ref{sec:DesignSensitivity}.
The radial angular acceptance $\Omega_R$, as seen from the trap centre, is 
$35.9^\circ$. It is limited by the axial position of
rings $R_3$ and $R_4$. The axial angular acceptance $\Omega_A$ is $55.4^\circ$ 
and is limited by the inner radius of ring 
$R_1$ or $R_2$. This acceptance is large enough to install the two annular 
detectors (with a hole radius of 6~mm)
before and after the Einzel lens triplets at 70~mm from the trap 
centre~\cite{Delahaye:2018}.

\begin{figure}
\centering
\includegraphics[width=0.95\linewidth]{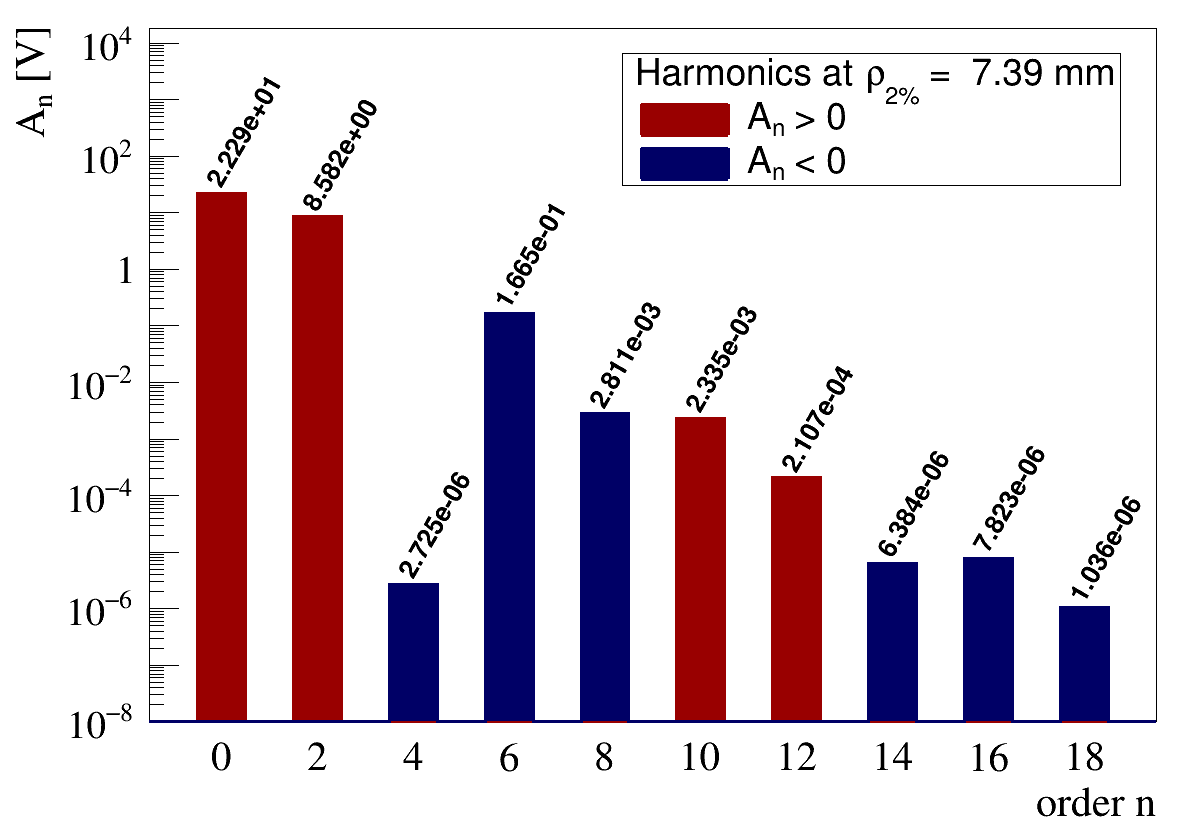}
\caption{ Harmonics coefficients for \morat up to order 18, extracted from potential sampled on a circle of 
radius $\rho_{2\%} = 7.39$~mm using {\textsc{axielectrobem}}, when applying 60~V on the innermost 
electrodes and 0~V on all the other electrodes. Positive and negative harmonics have been separated in order to display them with logarithmic scale.}
\label{fig:harmMORA}
\end{figure}

\begin{figure}
\centering
\includegraphics[width=0.90\linewidth]{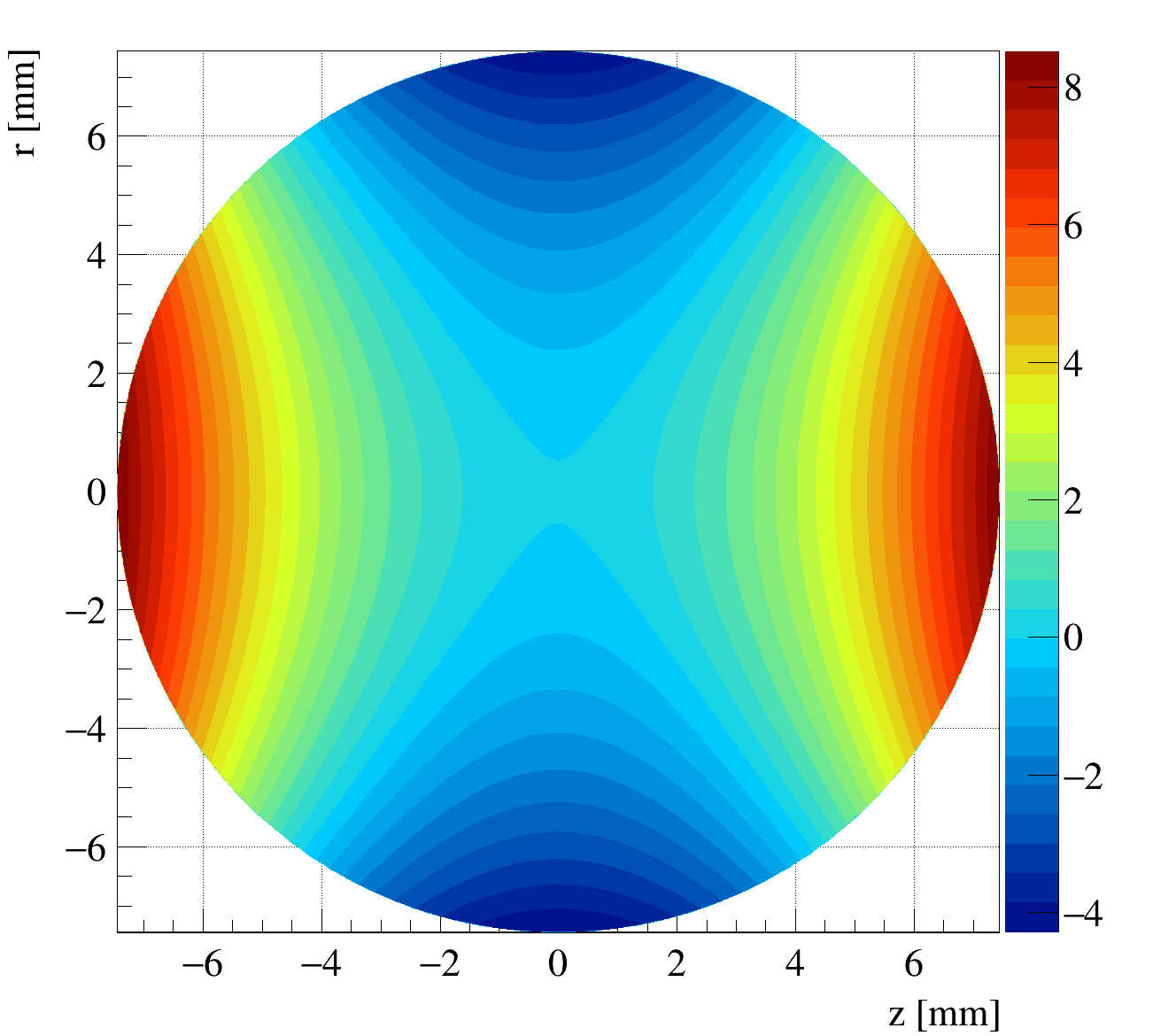} \\
\includegraphics[width=0.90\linewidth]{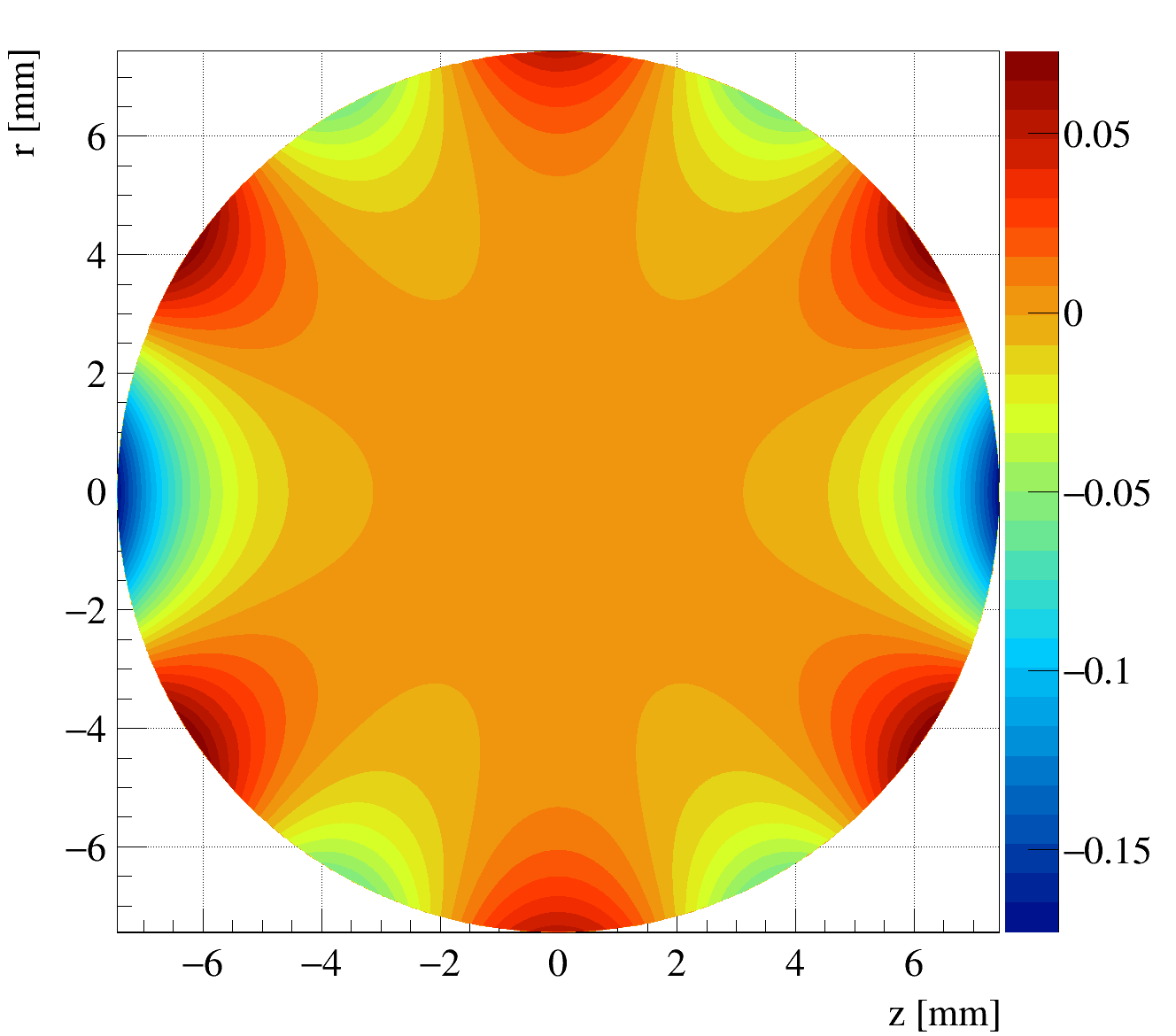} \\
\includegraphics[width=0.90\linewidth]{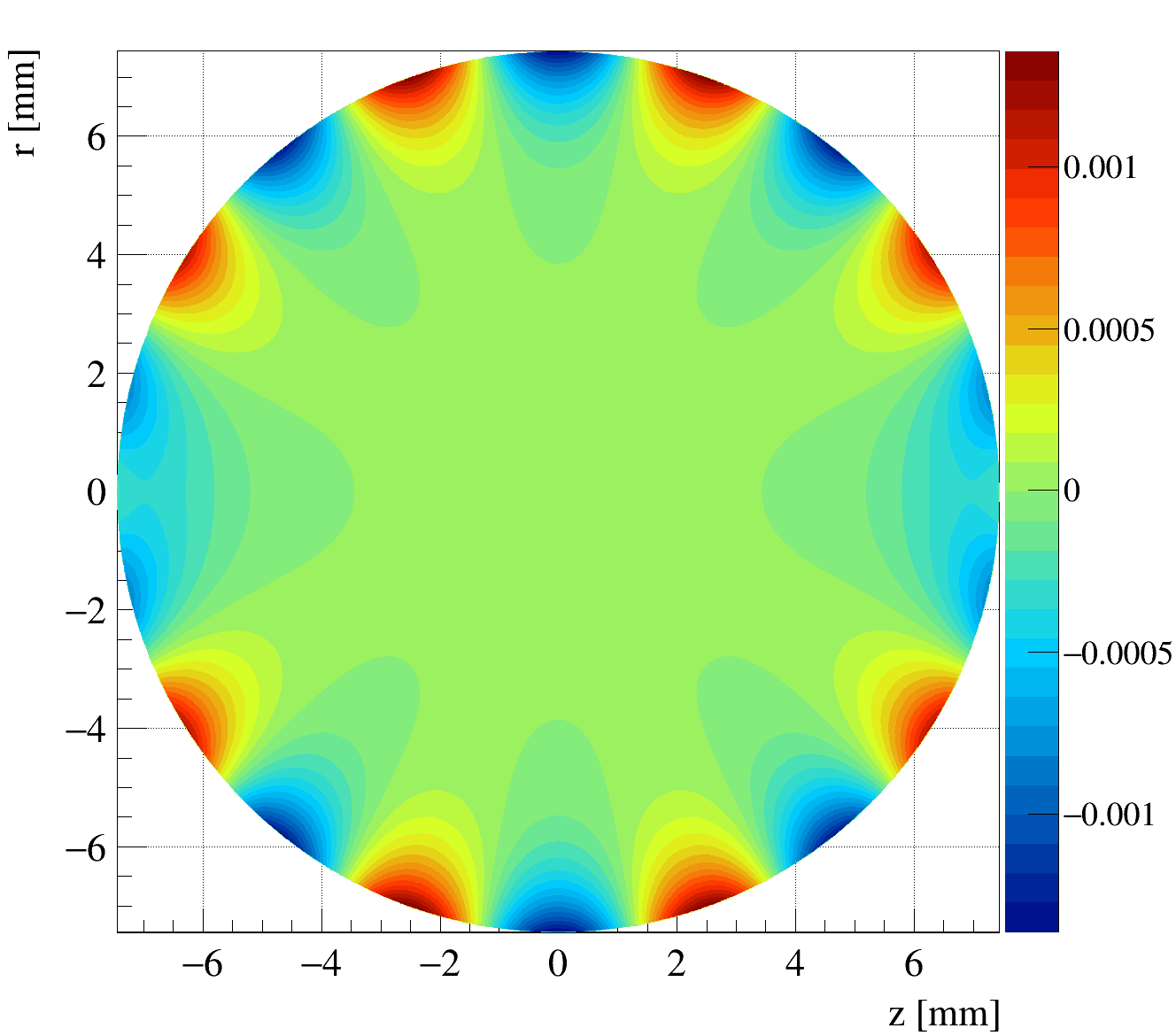}
\caption{ Contribution to the potential for $\rho \leqslant 7.39$~mm  in \morat from different harmonics sets (vertical scale is in volts). Top panel corresponds to harmonics with $n \geqslant 2$ and exhibits a strong quadrupole dependence. Middle panel is for $n \geqslant 4$ and clearly shows the dodecapole contribution. Bottom picture shows the contribution from all harmonics with $n \geqslant 8$.}
\label{fig:contrib}
\end{figure}

The harmonic coefficients $A_{n}$ ($n\in[0, 18]$) of the potential, computed at $\rho_{2\%}$, 
are presented in Fig.~\ref{fig:harmMORA}. The ratios $\frac{A_4}{A_2}$,  $\frac{A_6}{A_2}$  and 
$\frac{A_8}{A_2}$ are respectively around $-3.2\times10^{-7}, \,-1.9 \times10^{-2}$ 
and $-3.3 \times10^{-4}$, they confirm the dominance of the quadrupole term ${A_2}$.
Top panel in Fig.~\ref{fig:contrib} shows the potential 
(without the constant term $n = 0$) in the trapping region of \morat for $\rho \leqslant \rho_{2\%}$.  
Middle and bottom panels demonstrate how small the contribution of these higher multipoles ($n > 2$) is.
Since the octupole term $A_4$ is of the order of $10^{-6}$, the largest contribution comes from the dodecapole 
term $A_6$. Bottom panel of Fig.~\ref{fig:contrib} presents in more details the contribution from harmonics with orders greater or equal to 8: they are even lowered by about a factor 10 to 100 depending on the angular position.
These results demonstrate that the minimisation procedure led to a quadrupole potential with the 
required quality, a relative difference to $A_2$ lower than 2\% (see Eq.~(\ref{eq:2pc})), in a region  with a radius as large as 7.39~mm.
  
In next section we shall compare these performances to those of \lpct.

%
%
\section{Comparison of MORATrap and LPCTrap}
\label{sec:Comparison}

\begin{figure}[hbt!]
\centering
\includegraphics[width=0.95\linewidth]{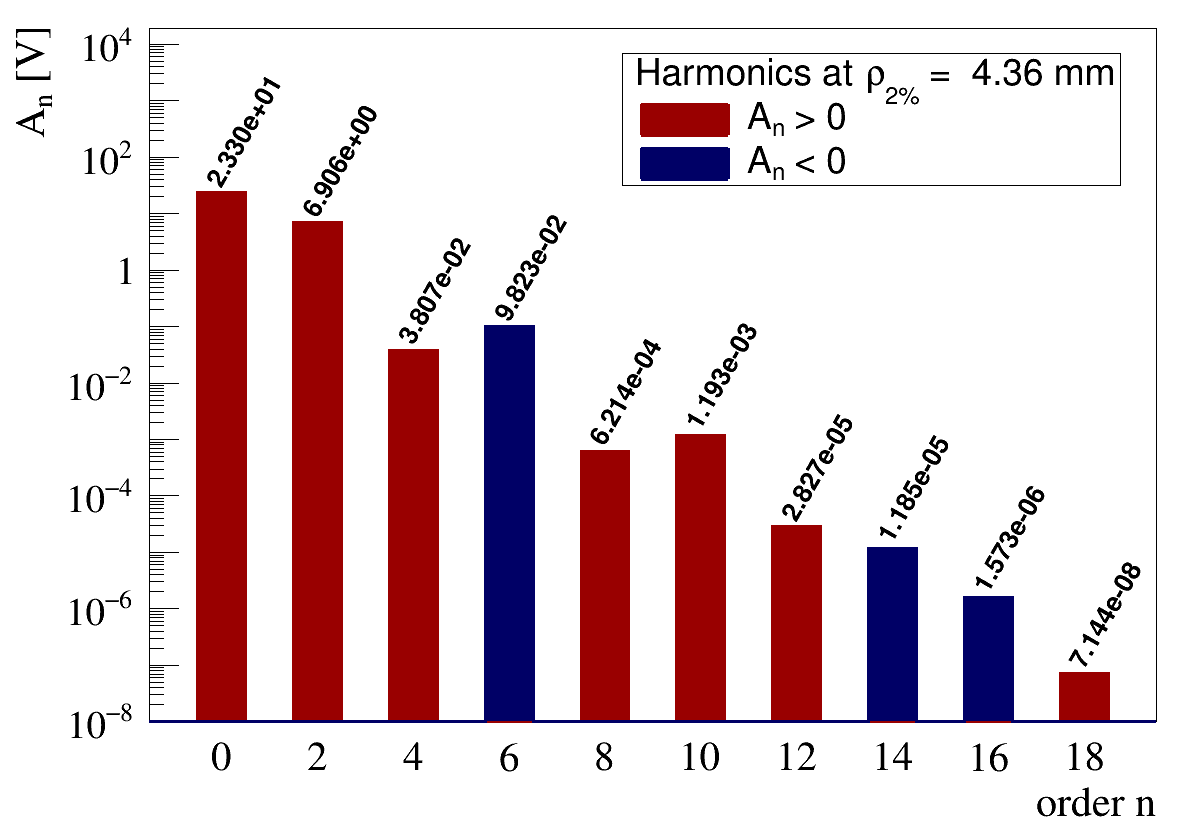}
\caption{Harmonics coefficients for \lpct up to order 18, extracted  from potential sampled on a circle of 
radius $\rho_{2\%} = 4.36$~mm using {\textsc{axielectrobem}}, when applying a 60~V RF voltage on the 
innermost electrodes.}
\label{fig:harmLPC}
\end{figure}

It is interesting to compare \morat performances with those of \lpct. For this purpose, 
the potential generated in \lpct (Fig.~\ref{fig:xsv_LPCTrap}) was simulated in {\textsc{axielectrobem}}
and the harmonics were determined as seen in Fig.~\ref{fig:harmLPC},  where obviously the octupole
term is much larger than in \morat. One should however notice that $A_4$ and $A_6$ have opposite signs,
thus partly compensating each others in the low radii region before $A_6$ takes over at larger radii. 
The ratios $\frac{A_4}{A_2}$,  $\frac{A_6}{A_2}$  and $\frac{A_8}{A_2}$ are respectively around 
$+5.5\times10^{-3}, \,-1.4 \times10^{-2}$ and $+9.0 \times10^{-5}$.
The corresponding radius $\rho_{2\%}$  is equal to 4.36~mm which is about 40\% smaller than 
for \morat. This clearly demonstrates that we have succeeded to broaden the 
trapping region in \morat and to improve the potential quality  compared to \lpct, given some maximal constraints in dimension for the Paul trap. In addition, due to 
its tube-shape electrodes, \lpct has an axial angular aperture of $43.6^\circ$ whereas for \morat
$\Omega_A = 55.4^\circ$, {\it i.e.} 27\% larger. The radial angular acceptances $\Omega_R$ are similar
 in both traps,
respectively $38.0^\circ$ and $35.9^\circ$ with a small advantage of 6\% for \lpct. However, it is
important to emphasise that in a trap, the amount of particles which can be effectively trapped
is directly proportional to the depth of the potential well {\it i.e.} the value of $A_2$ and to the r
adius $\rho_{2\%}$ (Eq.~(\ref{eq:Qmax})). One can note that in \morat, 
at $\rho = 4.36$~mm, to reach the same value of the quadrupole term as in \lpct, would require to apply 
138~V instead of 60~V like in the simulations presented throughout this paper.

To deepen this comparison it is interesting to make a parallel with an ideal Paul trap, introduced 
in Sec.~\ref{sec:Introduction}. For real Paul traps, one can define an equivalent internal radius 
$r_0$ as defined for ideal Paul traps in Eq.~(\ref{eq:Videal}), representing the radius at which 
the RF voltage would be applied to yield the same $A_2$ value at a given $\rho_{2\%}$:
\begin{equation}
r_0^2 \, = \, \rho^2_{2\%} \,\frac{V_0}{A_2}.
\label{eq:r0A2}
\end{equation}
Compared to an ideal Paul Trap, \morat has an effective internal radius $r_0$
significantly larger than the one of \lpct which explains the relative
difference in $A_2$ values at a given radius for both traps.  
From the $A_2$, $\rho_{2\%}$ and $V_0$ values shown in Figs.\ref{fig:harmMORA} and 
\ref{fig:harmLPC}, one finds an effective internal radius of 19.24~mm for \morat, and of 12.86~mm for 
\lpct. This is consistent with what was found in \cite{Delahaye:2019} and is summarised in 
Table~\ref{tab:comparison}. It is important however to note  that simply scaling \lpct to the same 
internal radius as \morat would not have
yielded the same improvements: the 2\% radius would have been enlarged to 
only 6.53~mm, compared to 7.39~mm in the case of \morat, and the axial angular acceptance
$\Omega_A$ would still have been limited by 27\% compared to \morat.
The trap capacity of \morat, estimated thanks to Eq.~(\ref{eq:Qmax}) and considering 
$r_{\mathrm{eff}} \, \simeq \, \rho_{2\%}$, has been enlarged by more than a factor of 2 
compared to the original \lpct, and by 40\% compared to a scaled version of \lpct.


\begin{table}[hbt!]
\ra{1.2}
\begin{center}
\begin{tabular}{@{}llll@{}}
\toprule
 & \bfseries{\lpct} & \bfseries{\morat} & \bfseries{\lpct} \\
 &                  &                   & \bfseries{scaled} \\
\midrule[0.25pt]
$V_0$ [V] & 60 & 60 & 60 \\
$r_0$ [mm] & 12.86 & 19.24 & 19.24 \\
$\rho_{2\%}$ [mm] & 4.36 & 7.39 & 6.53 \\
$A_2(\rho_{2\%})$ [V] & 6.91 & 8.58 & 6.91 \\
$\rho_{2\%} A_2$ [V.mm] & 30.14 & 63.44 & 45.09 \\
Maximum capacity & \num{1.05e6} & \num{2.20e6} & \num{1.56e6} \\
\bottomrule
\end{tabular}
\end{center}
\caption{Comparison of \morat and \lpct when both traps are assumed to behave like ideal Paul traps 
exhibiting a pure quadrupole potential. The maximum capacity was estimated for a pseudo 
potential model with a Mathieu parameter $q_z = 0.4$ and for singly charged positive ions. 
After rescaling \lpct effective internal radius to the $r_0$ of \morat, its radius $\rho_{2\%}$ 
is still smaller than the one obtained for \morat and its maximum capacity is about 30\% smaller.} 
\label{tab:comparison}
\end{table}

 \begin{figure}[hbt!]
\centering
\includegraphics[width=0.95\linewidth]{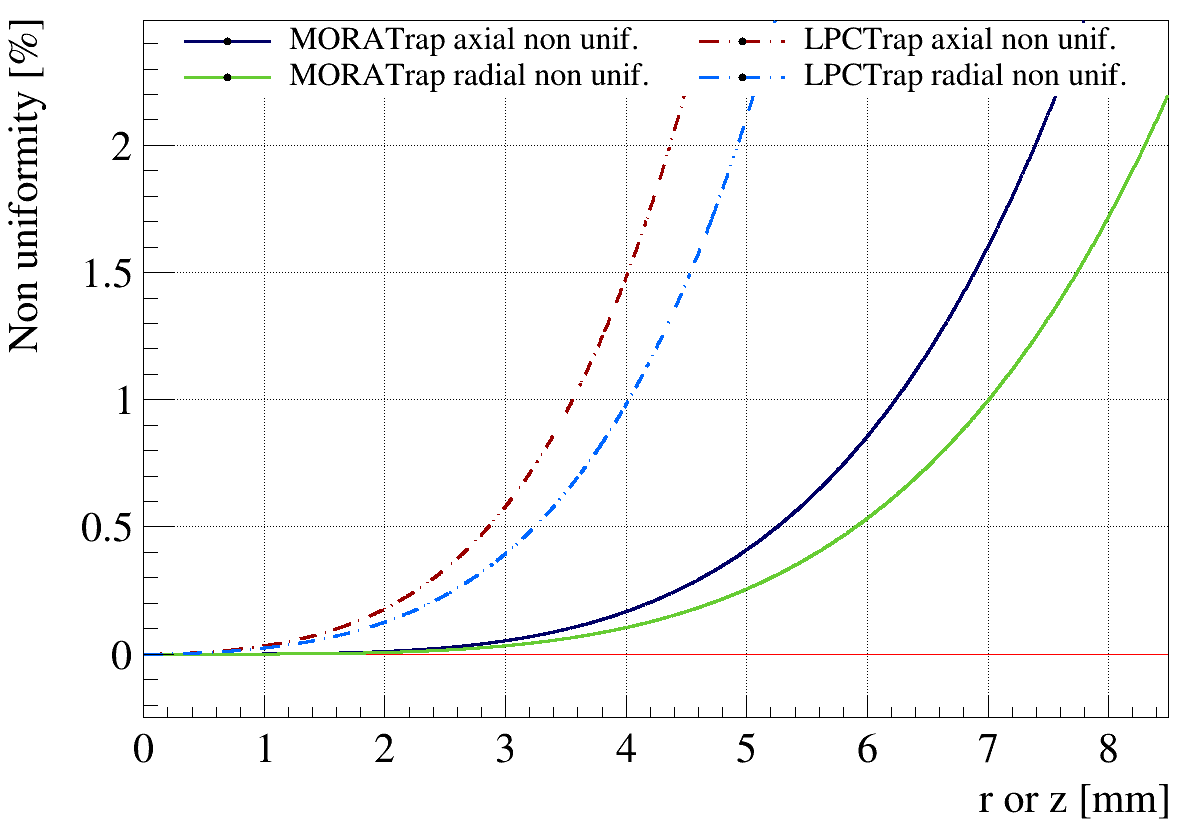}
\caption{Radial and axial non uniformities, respectively given by Eq.~(\ref{eq:unifr}) and Eq.~(\ref{eq:unifa}), 
as a function of the radial and axial distance from \morat centre.}
\label{figure8}
\end{figure}

To further compare both traps, one can investigate the potential non uniformity as a function of the space coordinates.
We can for instance define a radial and an axial non uniformity respectively by: 

\begin{eqnarray}
 U_{\mathrm{Radial}}(r) & = & \sum_{n=4}^{18} \left(\frac{r}{ R_0}\right)^{\!\!n-2} \left| \frac{A_n P_n(0)}{A_2 P_2(0)} \right| \,\,\,\, \mathrm{ at } \,\,\,\, z = 0,
   \label{eq:unifr} \\
  U_{\mathrm{Axial}}(z) & = & \sum_{n=4}^{18} \left(\frac{z}{ R_0}\right)^{\!\!n-2} \left| \frac{A_n}{A_2} \right| \,\,\,\, \mathrm{ at } \,\,\,\, r = 0,
   \label{eq:unifa} 
\end{eqnarray}
where the Legendre polynomial $P_n(\cos\theta)$ present in Eq. (\ref{eq:CH}) is equal to unity along $z$-axis and to $P_n(0)$ along $r$-axis. 
These quantities are presented in Fig.~\ref{figure8} where, in agreement with the values of $\rho_{2\%}$ found previously, the region where the radial and axial non uniformities are very weak is wider in the case of \morat.

 \begin{figure}[hbt!]
    \centering \includegraphics[width=0.90\linewidth]{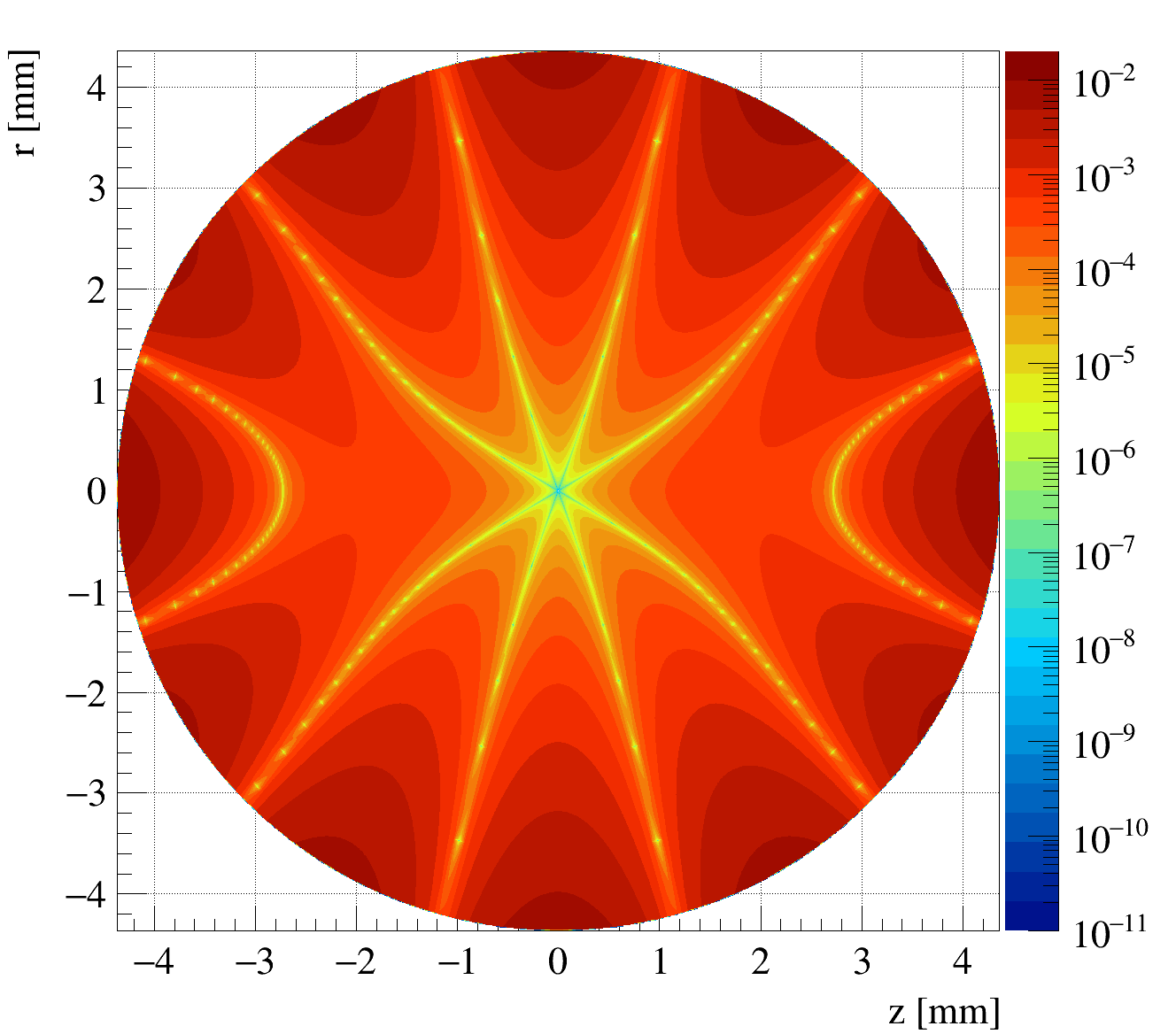}
    \centering \includegraphics[width=0.90\linewidth]{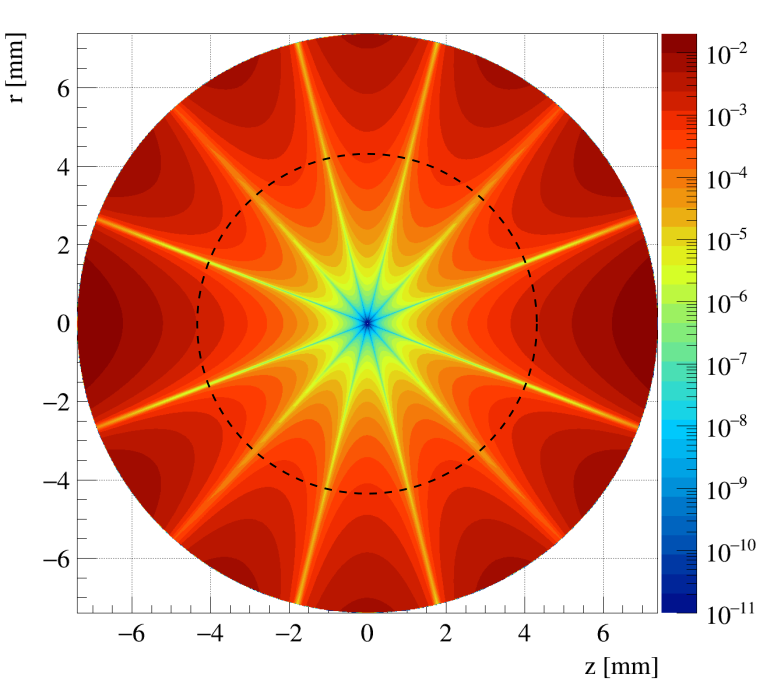} \\
    \caption{{\bfseries{Top:}} Potential non uniformity obtained with Eq.~(\ref{eq:unifar}), for \lpct region of
    interest, {\it i.e.} for $\rho = \sqrt{r^2 + z^2} \leqslant \rho_{2\%} = 4.36$~mm. {\bfseries{Bottom:}} 
    Same figure for \morat with $\rho_{2\%} = 7.39$~mm. The dashed black circle corresponds to the \lpct 2\% radius.}
    \label{fig:2Dunif}
\end{figure}

Another way to visualise the difference between the two traps is to draw a 2D uniformity defined as:
\begin{equation}
   U(r,z) =  \left|\sum_{n=4}^{18} \left(\frac{r}{ R_0}\right)^{\!\!n-2}  \frac{A_n}{A_2} \, P_n(\cos\theta)\right|.
   \label{eq:unifar}
\end{equation}  
where the absolute value has been chosen such that a logarithmic scale may be used to help distinguishing
more details. Figure~\ref{fig:2Dunif} shows $U(r, z)$ for $\rho \leqslant 4.36$~mm in the case of 
\lpct (top panel) and $\rho \leqslant 7.39$~mm in the case of \morat (bottom panel). The dashed black circle on bottom panel represents the limit of the region where the potential non uniformity is smaller than 2\% in \lpct.

All these different definitions of the non uniformity conclude to the improvement of the potential quality and 
of the trapping region volume in \morat compared to \lpct.

In next section, we shall investigate how stable is the potential against misalignment and other mechanical 
characteristics.

%
%
\section{Design sensitivity} 
\label{sec:DesignSensitivity}

\begin{figure}
\centering
\includegraphics[width=0.95\linewidth]{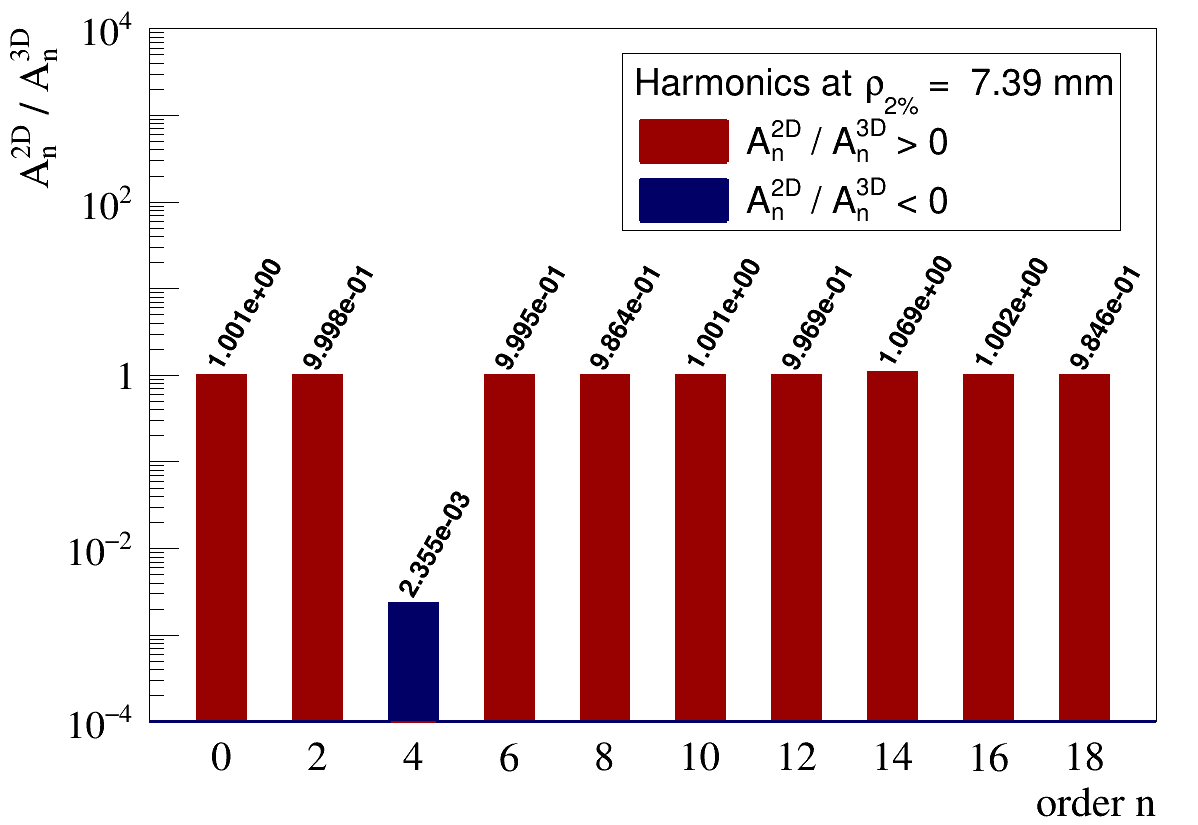}
\caption{Comparison between harmonics computed with {\textsc{axielectrobem}} and {\textsc{electrobem}} with an 
angular/polygonal segmentation of 84 cells around $z$-axis. Except for $A_4$, all harmonics are equal to better than a few percents, especially for low order ones.}
\label{figureAXEL}
\end{figure}  

It is important to check how the mechanical tolerances in the machining and assembling of the different electrodes of \morat can affect the trapping potential.
For this purpose, the effects on the potential due to mechanical precision, in terms of machining 
defects or misalignment of one or several electrodes, were investigated using both {\textsc{axielectrobem}} 
(2D) and {\textsc{electrobem}} (3D). {\textsc{axielectrobem}} was preferred in case of defects not breaking
axisymmetry, whereas {\textsc{electrobem}} was used in other cases. When performing simulations within 
{\textsc{electrobem}}, the setup needs to be meshed not only in the axial direction, but also angularly around 
this axis: a circle is approximated by a polygon and thus any cone, disk or cylinder is segmented in many
triangular or quadrangular cells.
As mentioned in Sec.~\ref{sec:LaplaceSolver}, the computer RAM usage rapidly increases when working in 3D compared to 
axially symmetric simulations. On the computer used, the available RAM allowed us to divide the setup in 84 cells 
in angle while keeping the same axial segmentation as the one used in {\textsc{axielectrobem}} leading to 
84 $\times$ 414 = 34776 cells in total. Such a meshing has consequences on the computed  harmonics spectrum. 
Figure~\ref{figureAXEL} shows the ratio of the harmonics from the 2D axisymmetric case to the 3D one. There is 
a very good agreement between the two: it is better than a few percents 
except for the octupole term $A_4$, which appears to be very sensitive to the angular discretisation and dramatically changes not only in absolute value, but also in sign. Nevertheless, the 3D computation leads to a
value of $A_4$ still about a factor 100 less than $A_6$ and its contribution to the potential is therefore very limited. When using 3 planes 
of symmetry in {\textsc{electrobem}}, one can simulate 1/8 of \morat with as many cells as a full simulation 
without symmetry planes.
In that case, the 3D harmonics spectrum agrees with the 2D one within a few $10^{-6}-10^{-5}$ even for $A_4$. However, 
most of the envisaged defects break at least one of these symmetries. Consequently, the full \morat has to be 
simulated in these studies and the results should be compared to those from the perfectly aligned 3D version 
of \morat rather than to those from {\textsc{axielectrobem}}.
The precision on the evaluated potential essentially depends on the characteristics of the electrode mesh 
and is constant once this mesh has been fixed. A single mesh is used to test different trap configurations 
with or without misalignment. The results obtained for configurations exhibiting some defects are compared 
to a reference one where the electrodes are perfectly aligned and machined (in the limit of the mesh
discretisation) and therefore one expects that the observed differences are highly significant overall.

The considered defects included machining 
tolerances of $\delta r$ and $\delta z \leqslant \SI{20}{\micro\metre}$, axial and radial translations of $\Delta x$ and 
$\Delta z \leqslant \SI{200}{\micro\metre}$ and rotations around $r$-axis of $\Delta\theta \leqslant~0.2^\circ$. 
All these values are larger than what can be mechanically achieved for \morat. These
defects were simulated for either one half of the trap in a whole 
 or for electrodes taken individually while keeping the rest of the trap unchanged. 

When dealing with individual electrodes, because of their relatively large distance from the trap ROI,
all types of defects related to $R_6, E_4, E_5$ and $E_6$ have been shown to have very minor influence
on the harmonics spectrum and therefore on the potential quality. This was also the case for possible
machining defects on $R_1, R_2, R_3$ and $R_4$ as the computer numerical control machining precision was assumed better than \SI{20}{\micro\metre}.

\begin{figure}[hbt!]
  \centering
  \includegraphics[width=0.95\linewidth]{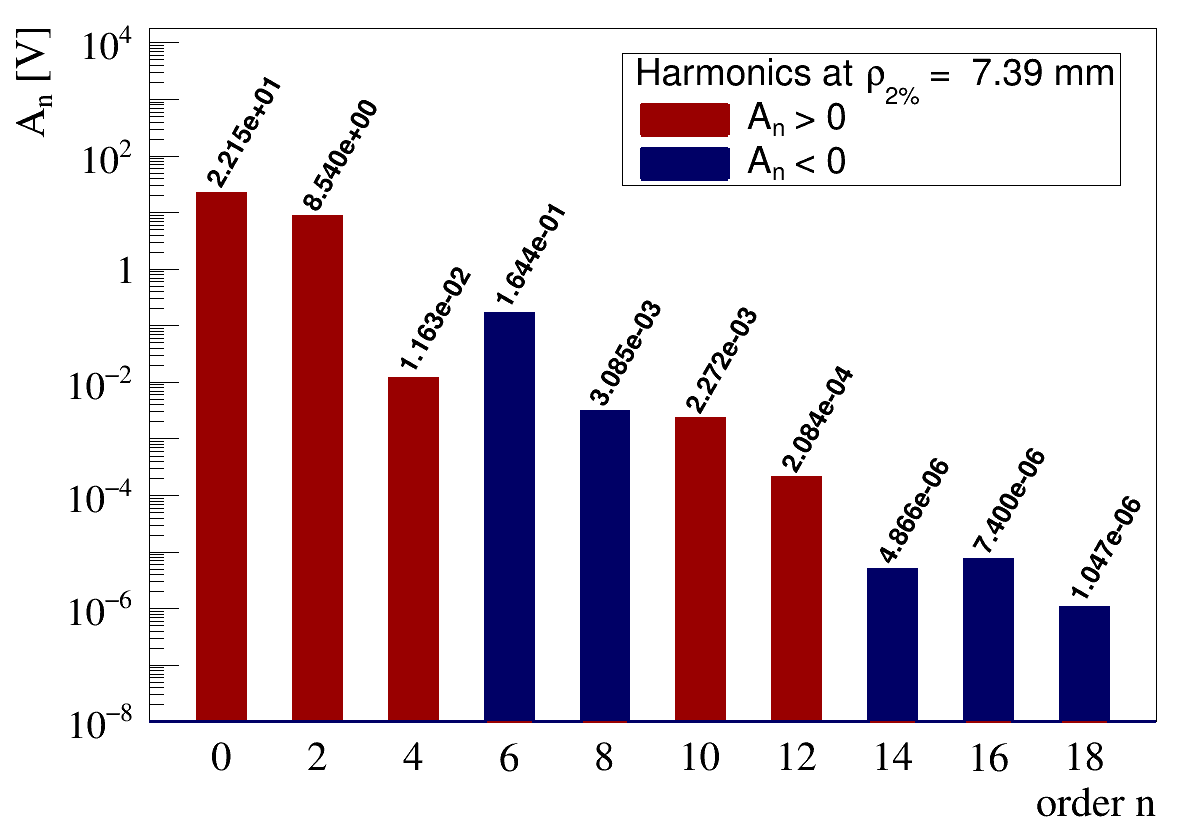}
  \caption{Harmonics spectrum obtained when translating left half of \morat by $\Delta z = -\SI{100}{\micro\metre}$.
  The expansion has been computed around a centre located at half of the displacement ($z = -\SI{50}{\micro\metre}$)  where 
  the translated potential centre stands, {\it i.e.} where the ion cloud barycentre will be.}
\label{fig:HT0p1mm_displacement}
\end{figure}

\begin{figure}[hbt!]
  \centering
  \includegraphics[width=0.90\linewidth]{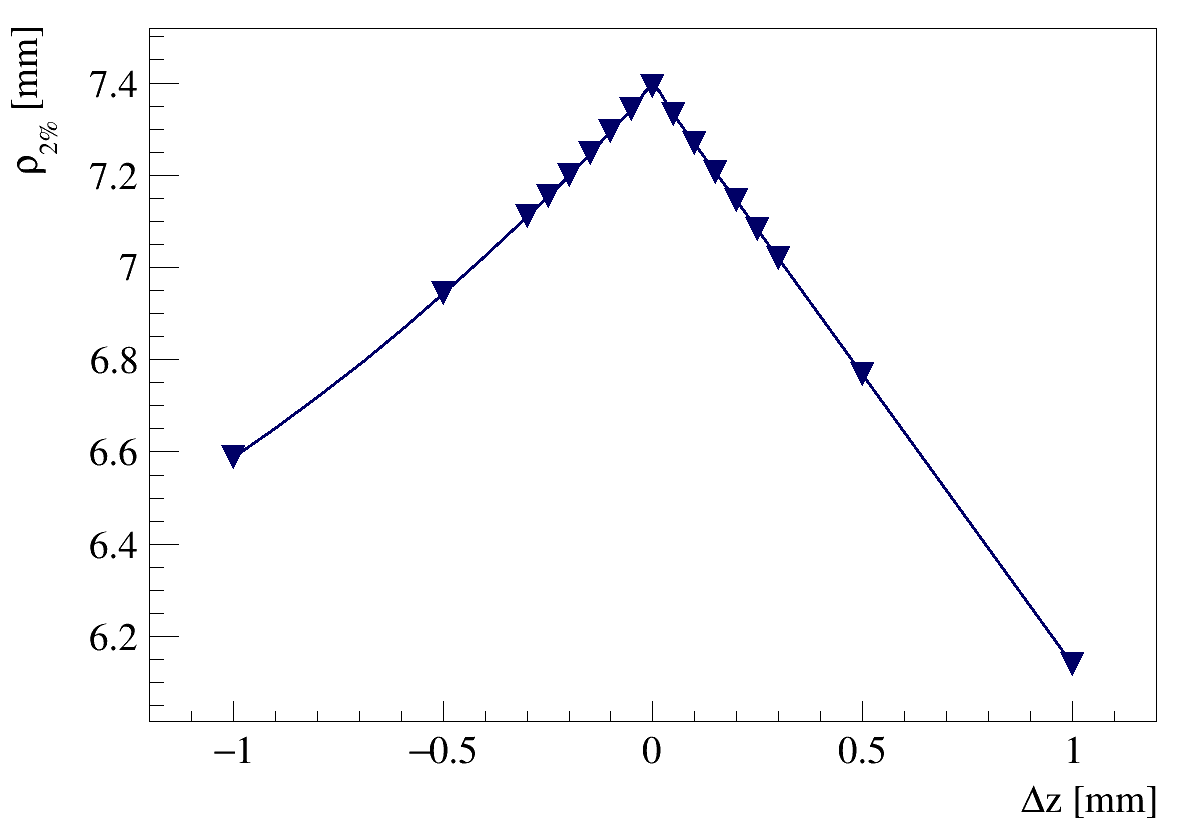} 
  \includegraphics[width=0.90\linewidth]{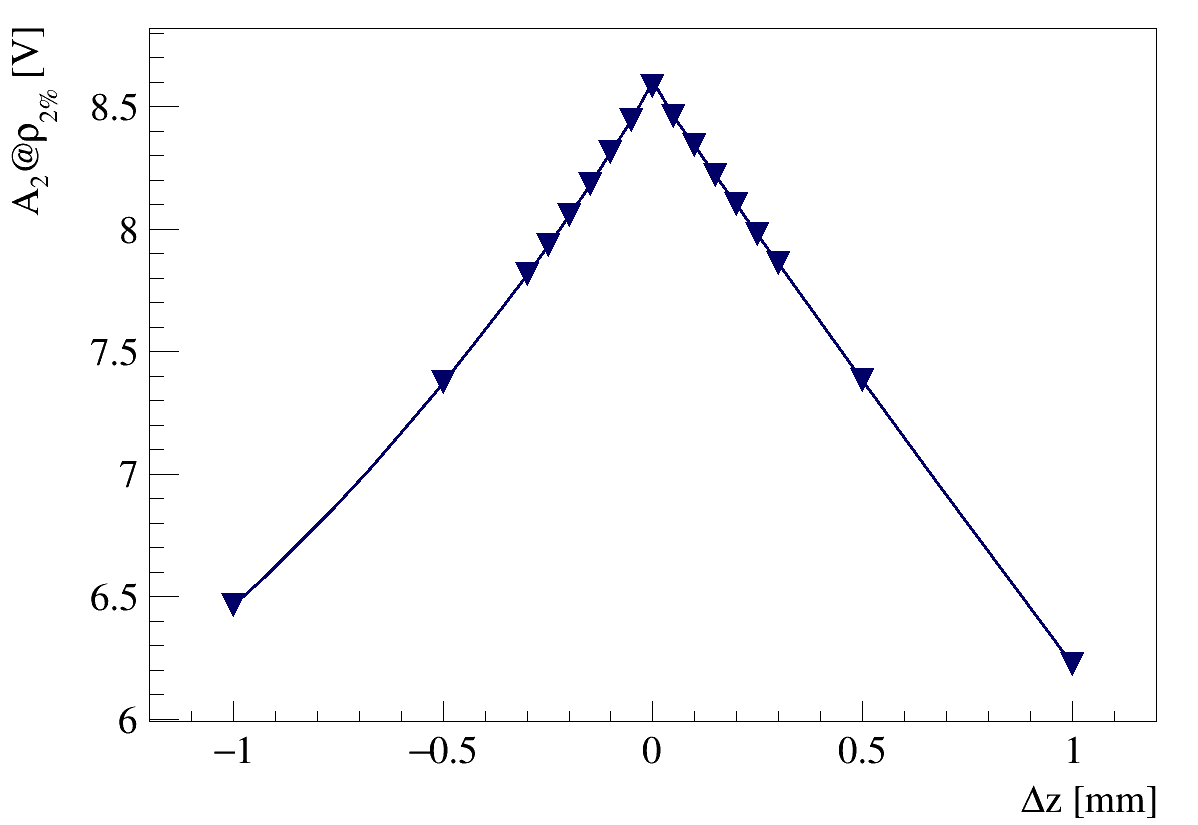}
  \includegraphics[width=0.90\linewidth]{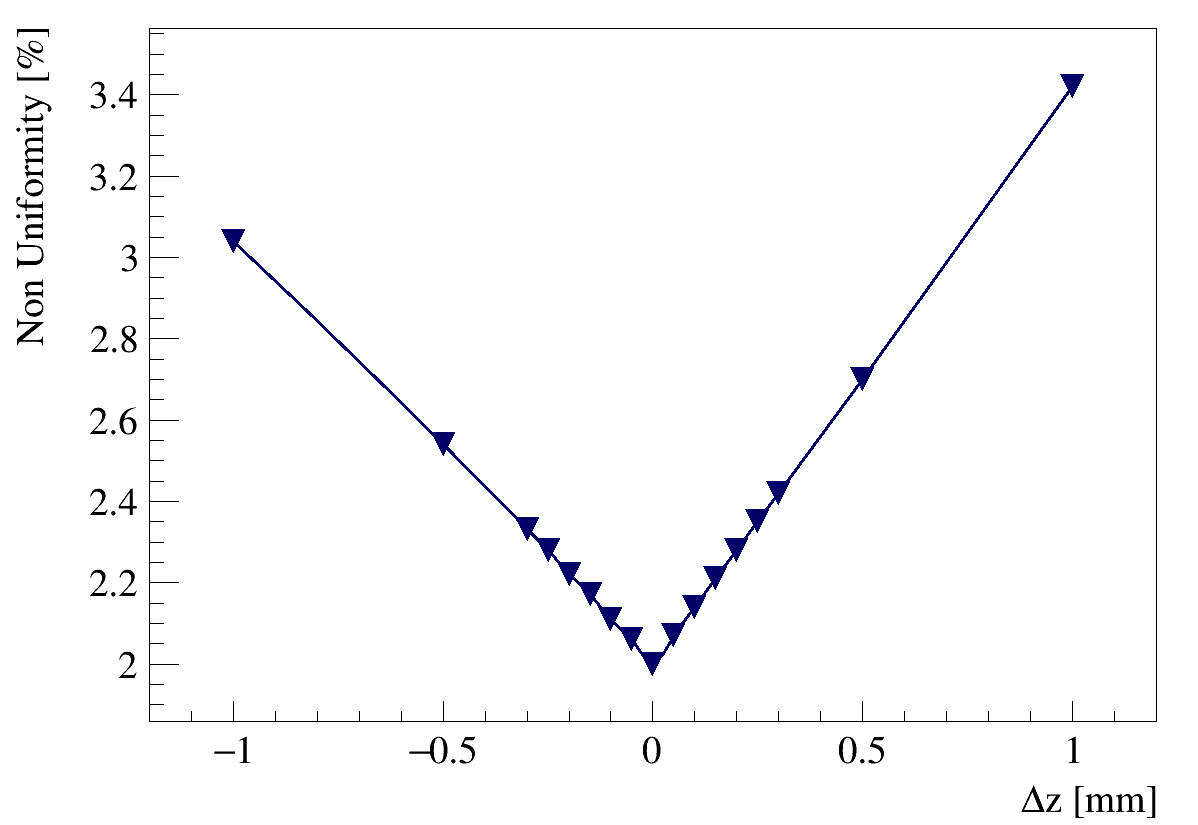}
  \caption{Evolution of the radius $\rho_{2\%}$, the quadrupole term $A_2 \, \mathrm{at} \, \rho_{2\%}$ and  the uniformity as a function of $\Delta z$, the translation distance of the left half of the trap 
  (at $z < 0$) from its nominal position. These results were obtained using {\textsc{axielectrobem}}.}
 \label{fig:deltaz}
\end{figure}

\begin{figure}[hbt!]
  \centering
  \includegraphics[width=0.90\linewidth]{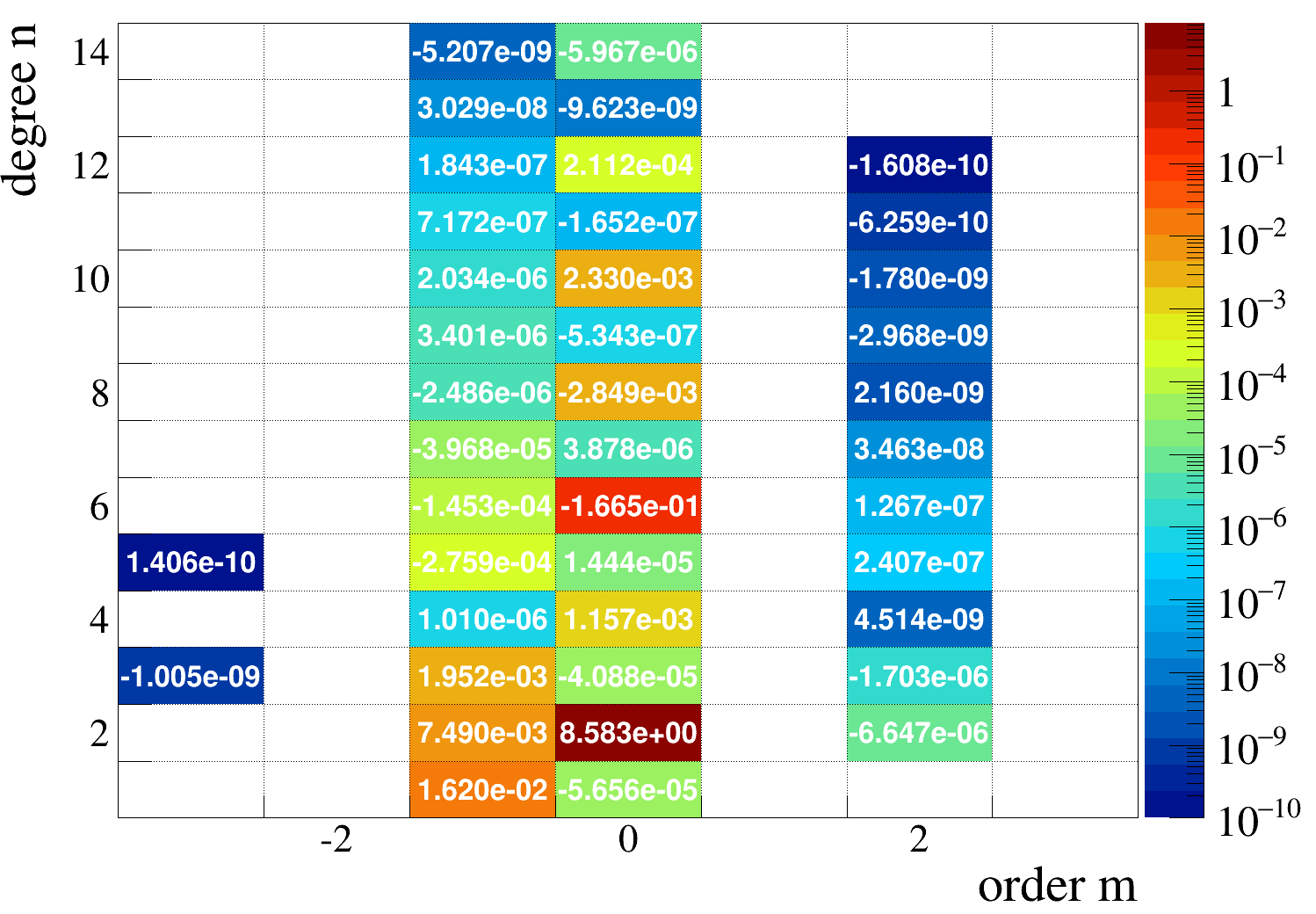}
  \caption{Harmonics spectrum (without the constant term $A_{0,0}$) obtained at $\rho_{2\%}$ with {\textsc{electrobem}} for half of the trap rotated by $0.2^\circ$ around $x$-axis (vertical scale is in volts). The series expansion centre is kept at the frame origin which does not correspond to the minimum of the potential after rotation. Harmonics with absolute values below $10^{-10}$ have been set to zero as their contribution to the potential is completely negligible.}
 \label{fig:rotX0p2deg}
\end{figure}

The influence of misalignment along the $z$ direction was studied with {\textsc{axielectrobem}} by varying
electrode(s) axial position. An axial displacement $\Delta z$  of half of the trap conserves both the axisymmetry and 
the planar symmetry: such a displacement shifts the trap centre by $\Delta z / 2$ 
and therefore the potential still exhibits a planar symmetry around the position of this shifted centre. 
The potential harmonics expansion should therefore be determined around this shifted centre to avoid
 inducing spurious 
harmonics with odd orders. In the particular case of ion traps, the ion cloud barycentre will 
follow the potential centre and, in case of a large shift, this may modify the acceptance and have some 
consequences 
both on the detector counting rates and on the measured $\beta$ decay asymmetry. This shall be addressed 
in other simulations.
Figure \ref{fig:HT0p1mm_displacement} presents the harmonics spectrum computed around a point located at 
$z = \Delta z / 2 = -\SI{50}{\micro\metre}$ for a global translation of all electrodes located at 
negative $z$ by 
$\Delta z= -\SI{100}{\micro\metre}$.  One clearly sees that odd harmonics are not present in this 
spectrum. In this case, the potential quality is still suitable to trap ions as may be seen 
in Fig.~\ref{fig:deltaz} which shows the evolution of $\rho_{2\%}$, $A_2$ and the non uniformity 
(Eq.~(\ref{eq:2pc}))
versus the translation $\Delta z$. A non uniformity of 2.5\% is reached for displacements 
as large as 0.4~mm and even for such large $\Delta z$, $\rho_{2\%}$ is still about 6.9~mm 
and $A_2$ is only lowered by less than 10\%.

To describe and further compare the harmonics induced by all the studied translations and rotations of different setup parts,
we shall use the spherical harmonics with degree $n$ 
and order $m$ (Eq.~(\ref{eq:SH})) instead of the cylindrical harmonics of order $n$ (Eq.~(\ref{eq:CH})) used so far. 
Indeed, breaking the axisymmetry and/or the planar symmetry may induce some harmonics with different degrees $n$ but also 
and above all with non zero order $m$ on the contrary to what has been seen previously. As an example, 
Fig.~\ref{fig:rotX0p2deg} shows the 
harmonics corresponding to a setup where half of the trap has been rotated by $\Delta\theta = 0.2^\circ$ around $x$-axis. 
One should notice that even if the potential centre has probably been shifted away from the reference frame origin, the
harmonics expansion is still performed around this origin. The largest harmonics generated by the rotation are however a 
factor 10 smaller than the main ones and should therefore weakly impact the quality of the trapping potential.
To estimate their influence and similarly to what was done in Eqs.~(\ref{eq:2pc}),  
(\ref{eq:unifr}) and (\ref{eq:unifa}), we shall define a 3D non uniformity. This is more difficult as the associated 
Legendre functions of the first kind $P_n^m(\cos\theta)$ are not bounded to $[-1, 1]$ on the contrary to the Legendre polynomials 
$P_n(\cos\theta)$: their minimum (resp. maximum) values strongly decrease (resp. increase) with degree $n$ and order $m$.
However they always satisfy $\max(P_n^m(x)) \geqslant |\min(P_n^m(x))|$. This inspired us to define a 3D non uniformity 
at $\rho = R_0 = \rho_{2\%}$ (maximising the radial contribution) by:
\begin{equation}
 U_{\mathrm{3D}} \, = \, \sum_{n=1}^{n_{\mathrm{max}}} \sum_{\substack{m=0 \, \mathrm{if} \, n\neq 2\\m=1 \, \mathrm{if} \, n = 2}}^{n} \! \max(P_n^m(\cos\theta)) \frac{|A_{nm}| \, + \, |B_{nm}|}{|A_{2,0}|},
 \label{eq:u3d}
\end{equation}
where the numerator and therefore $U_{\mathrm{3D}}$ have been maximised by setting the functions $P_n^m(\cos\theta)$, 
$\cos(m\varphi)$ and $\sin(m\varphi)$ to their maximal values and by taking the absolute values of the harmonic coefficients.
For the denominator we only keep $|A_{2,0}|$ without any spatial dependent function which might become null at some location
leading to an infinite value of $U_{\mathrm{3D}}$.
When simulating \morat in {\textsc{electrobem}} without any misalignment, the 3D non uniformity $U_{\mathrm{3D}}$ is 2.017~\%,
which shall serve as our reference. One should notice that this value is weakly above the 2\% limit obtained with
{\textsc{axielectrobem}}, this highlights the influence of the 3D meshing.
The $U_{\mathrm{3D}}$ values obtained for the tested misalignment are presented in Tables~\ref{tab:deltazx} and \ref{tab:deltath} 
respectively for translations ($\Delta z$, $\Delta x$) and rotations ($\Delta\theta$). These clearly demonstrate
that  displacements of either rings $R_5$ and $R_6$ or Einzel lens triplets have negligible influence on the potential 
quality within the ROI. The largest non uniformities are obtained for translations $\Delta x$ of ring electrode $R_4$
or for the rotation of half of the trap. Even if these non uniformities were extracted from harmonics series expansions
performed around the reference frame origin instead of around the potential centre, they still remain smaller than 
2.5\%, {\it i.e.} at a level sufficient to maintain the trapping efficiency~\cite{Delahaye:2019}.

\begin{table}[t!]
\ra{1.2}
\begin{center}
\begin{tabular}{@{}lcccc@{}}
\toprule
\bfseries{Translated} & $\mathbf{\Delta z}$ & $\mathbf{U_{3D}}$ & $\mathbf{\Delta x}$ & $\mathbf{U_{3D}}$  \\
\bfseries{part} &  [\SI{}{\micro\metre}] & [\%] &  [\SI{}{\micro\metre}] & [\%] \\
\midrule[0.25pt]
Half trap & 100 & 2.127 & 200 & 2.398 \\  
$R_2$     & 100 & 2.182 & 200 & 2.208 \\
$R_4$     & 100 & 2.142 & 200 & 2.461 \\
$R_6$     & 100 & 2.017 & 200 & 2.017 \\
$E_4$     & 100 & 2.017 & 200 & 2.017 \\
$E_5$     & 100 & 2.017 & 200 & 2.017 \\
$E_6$     & 100 & 2.017 & 200 & 2.017 \\
\bottomrule
\end{tabular}
\end{center}
\caption{3D non uniformity due to $\Delta z$ and $\Delta x$ misalignment of the half trap and of individual electrodes.}
\label{tab:deltazx}
\end{table}
\begin{table}[t!]
\ra{1.2}
\begin{center}
\begin{tabular}{@{}lcc@{}}
\toprule
\bfseries{Rotated} & $\mathbf{\Delta}\bm{\theta}$ & $\mathbf{U_{3D}}$  \\
\bfseries{part} &  [$^\circ$] & [\%] \\
\midrule[0.25pt]
Half trap & 0.2 & 2.404 \\
$R_2$     & 0.2 & 2.267 \\
$R_4$     & 0.2 & 2.159 \\
$R_6$     & 0.2 & 2.018 \\
$E_4$     & 0.2 & 2.017 \\
$E_5$     & 0.2 & 2.017 \\
$E_6$     & 0.2 & 2.017 \\
\bottomrule
\end{tabular}
\end{center}
\caption{3D non uniformity due to $\Delta\theta$ misalignment. Here the series expansion centre is kept at the
frame origin. Realistic mechanical rotations of at most 0.2$^\circ$ lead to non uniformities well below 2.5\%. }
\label{tab:deltath}
\end{table}

\begin{figure}[hbt!]
  \centering
  \includegraphics[width=0.90\linewidth]{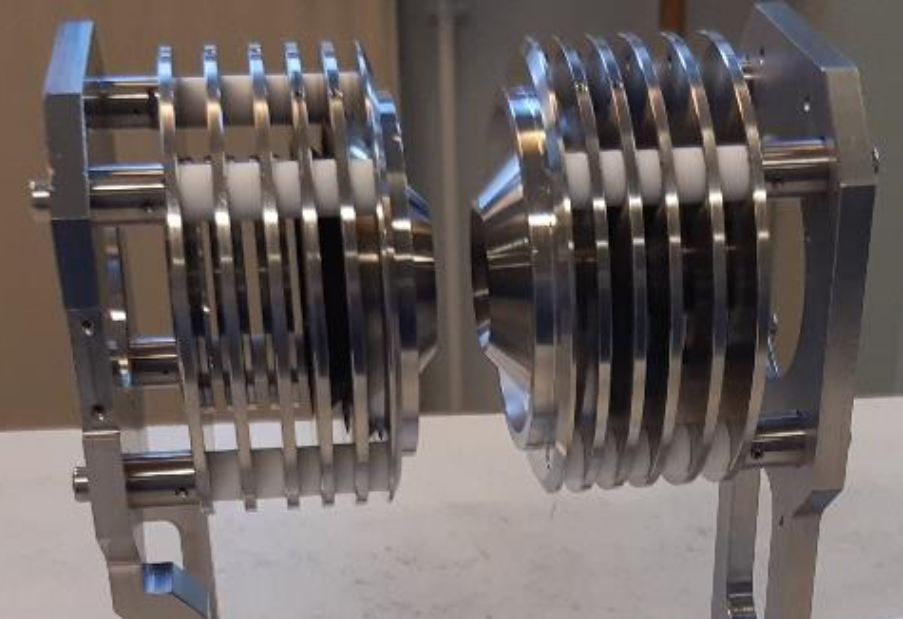}
  \caption{Picture of \morat on its mechanical support. It has been machined and assembled at LPC Caen.}
 \label{fig:MORATrap_picture}
\end{figure}

All the reasonable mechanical defects studied have been shown to weakly affect the quality of the potential 
and the volume of the trapping region in \morat if the machining process and the 
alignment of the different electrodes during assembly are kept under control. It was therefore decided to design, 
machine and assemble the \morat electrodes at LPC Caen. A picture of this realisation is shown in 
Fig.~\ref{fig:MORATrap_picture}.

%
%
\section{Conclusion}
\label{sec:Conclusion}

We have presented a method used to optimise the geometry of the axially symmetric ion trap, 
\morat, dedicated to the measurement of the triple correlation parameter $D$ in nuclear 
$\beta$-decay of  radioactive ions. Our starting point was the \lpct geometry, the former used
 transparent Paul trap, and we succeeded in
reducing the contribution to the potential from harmonics of order higher than 2. This is necessary 
to achieve a longer storage time and a better trapping efficiency.  According to the pseudo-potential
approximation, the \morat capacity has been enlarged by more than a factor of 2 compared to the 
initial \lpct geometry. 
The optimised geometry exhibits 
a larger axial angular acceptance than in \lpct. This larger
acceptance is necessary for the online monitoring of the trapped ion cloud polarisation, by detecting 
$\beta$ particles along the axis of the trap. Further
 simulations are currently being performed to deeply investigate the ion cloud dynamics, the 
 trapping efficiency as well as the trapping time. 
The optimised trap electrodes could easily be machined with a mechanical precision of the order 
of \SI{10}{\micro\metre}. 
The whole setup should be soon tested and then installed at JYFL accelerator facility in University of 
Jyv\"askyl\"a (Finland) before its final operation at GANIL - DESIR (Caen, France).

The optimisation method described in this study could easily be applied to other axisymmetric ion traps or 
electrodes setups. Different types of fitness functions could bring many improvements to the method and should be 
explored in detail.

%
%
\begin{acknowledgement}
This work was financially supported by R\'egion Normandie  via its R\'eseaux d'Int\'er\^ets Normands. 
The authors would like to thank their collaborators from the LPC Caen CAD group and workshop for their 
deep involvement in the design, manufacture and assembly of \morat.
\end{acknowledgement}

%
%
\begin{appendix}

%
%
\section{Homogeneous polynomials in cylindrical axisymmetric coordinates}
\label{app:HomogPoly}

For completeness, in this appendix, we present a series expansion for the potential of an axisymmetric 
system in cylindrical coordinates $(r, z)$. Starting from the coordinates conversion relations 
$\rho^2 = r^2 + z^2$ and $\cos\theta = z / \rho$ and introducing the homogeneous harmonic polynomials 
$H_n(r, z)$ given by an explicit relation rather than by a recurrence one:

\begin{eqnarray}
 H_n(r, z) & = & \rho^n \, P_n(\cos\theta) \nonumber \\
           & = & \sum_{k = 0}^{E[ n / 2]} \, \frac{n!}{(-4)^{k} (n - 2k)! (k!)^2} \, 
r^{2k} \, z^{n - 2k},
\end{eqnarray}
the potential given by Eq. (\ref{eq:CH}) is rewritten as :
\begin{equation}
 V(r, z) \, = \, \sum_{n = 0}^{\infty} \, \frac{A_n}{\rho_0^n} \, H_n(r, z).
 \label{eq:Hn}
\end{equation}
This series expansion converges for points $(r, z)$ satisfying $\sqrt{r^2 + z^2} \leqslant \rho_0$ with $\rho_0$
the convergence radius.
Again, a planar symmetry at $z = 0$ allows to further reduce the multipole expansion to even 
harmonics only. The first harmonic polynomials are listed up to order 10 in Table \ref{tab:Hn}.

\begin{table}[h]
\centering
\ra{1.2}
\begin{tabular}{@{}rl@{}}
\toprule
$\mathbf{n}$ & $\mathbf{H_n(r, z)}$ \\
\midrule[0.25pt]
0   & $1$\\
1   & $z$ \\
2   & $\frac{1}{2} \left(2 z^2-r^2\right)$ \\
3   & $\frac{1}{2} \left(2 z^3-3 r^2 z\right)$ \\
4   & $\frac{1}{8} \left(3 r^4-24 z^2 r^2+8 z^4\right)$ \\
5   & $\frac{1}{8} \left(8 z^5-40 r^2 z^3+15 r^4 z\right)$ \\
6   & $\frac{1}{16} \left(-5 r^6+90 z^2 r^4-120 z^4 r^2+16 z^6\right)$ \\
7   & $\frac{1}{16} \left(16 z^7-168 r^2 z^5+210 r^4 z^3-35 r^6 z\right)$ \\
8   & $\frac{1}{128} \left(35 r^8-1120 z^2 r^6+3360 z^4 r^4-1792 z^6 r^2+128 z^8\right)$ \\
9   & $\frac{1}{128} \left(128 z^9-2304 r^2 z^7+6048 r^4 z^5-3360 r^6 z^3\right.$ \\
    & $\qquad\left.+315 r^8 z\right)$ \\
10  & $\frac{1}{256} \left(-63 r^{10}+3150 z^2 r^8-16800 z^4 r^6+20160 z^6 r^4\right.$ \\
    & $\qquad\left.-5760 z^8 r^2+256 z^{10}\right)$ \\
\bottomrule
\end{tabular}
\caption{Harmonic polynomials of degree $n \leqslant 10$ in axisymmetric cylindrical coordinates $(r, z)$.}
\label{tab:Hn}
\end{table}

%
\end{appendix}

%
%
 \bibliographystyle{unsrt}
 \bibliography{MORA}
%
\end{document}